\documentclass[preprint,nofootinbib,amsmath,prd,aps,superscriptaddress,tightenlines,12pt]{revtex4}
\usepackage{graphicx}
\usepackage{graphics}
\usepackage{amsmath}
\usepackage{hyperref}
\usepackage{xcolor}
\usepackage{xspace}
\usepackage{subfigure}

\newcommand{\eq}[1]{Eq.~\eqref{eq:#1}}
\newcommand{\eqs}[2]{Eqs.~\eqref{eq:#1} and \eqref{eq:#2}}

\renewcommand{\sec}[1]{Sec.~\ref{sec:#1}}

\newcommand{\subsec}[1]{Sec.~\ref{subsec:#1}}

\newcommand{\fig}[1]{Fig.~\ref{fig:#1}}

\newcommand{\tab}[1]{Table~\ref{tab:#1}}

\def\lsim{\mathrel{\rlap{\lower4pt\hbox{\hskip1pt$\sim$}}
    \raise1pt\hbox{$<$}}}                
\def\gsim{\mathrel{\rlap{\lower4pt\hbox{\hskip1pt$\sim$}}
    \raise1pt\hbox{$>$}}}                

\def\OMIT#1{}

\newcommand{\be}{\begin{eqnarray}}
\newcommand{\ee}{\end{eqnarray}}

\newcommand{\nn}{\nonumber}

\newcommand{\bea}{\begin{eqnarray}}
\newcommand{\eea}{\end{eqnarray}}

\newcommand{\tot}{\textrm{tot}}
\newcommand{\res}{\textrm{res}}
\newcommand{\cut}{\textrm{cut}}
\newcommand{\FO}{\textrm{FO}}
\newcommand{\kt}{\textrm{k}_{\textrm{T}}}
\newcommand{\pTcut}{p_T^{\rm cut}}
\newcommand{\pTjet}{p_T^{\rm jet}}
\newcommand{\pToff}{p_T^{\rm off}}
\newcommand{\as}{\alpha_s}
\newcommand{\GeV}{\,\mathrm{GeV}}

\newcommand{\pb}{\,\mathrm{pb}}
\newcommand{\ord}[1]{\mathcal{O}(#1)}

\def\lsim{\mathrel{\rlap{\lower4pt\hbox{\hskip1pt$\sim$}}
    \raise1pt\hbox{$<$}}}                
\def\gsim{\mathrel{\rlap{\lower4pt\hbox{\hskip1pt$\sim$}}
    \raise1pt\hbox{$>$}}}                

\def\OMIT#1{}

\begin{document}

\setlength\baselineskip{17pt}

\begin{flushright}
\vbox{
\begin{tabular}{l}
DESY 13-244
\end{tabular}
}
\end{flushright}
\vspace{0.1cm}


\title{\bf Combining Resummed Higgs Predictions Across Jet Bins}

\vspace*{1cm}

\author{Radja Boughezal}
\email[]{rboughezal@anl.gov}
\affiliation{High Energy Physics Division, Argonne National Laboratory, Argonne, IL 60439, USA} 
\author{Xiaohui Liu}
\email[]{xiaohui.liu@northwestern.edu}
\affiliation{High Energy Physics Division, Argonne National Laboratory, Argonne, IL 60439, USA} 
\affiliation{Department of Physics \& Astronomy, Northwestern University, Evanston, IL 60208, USA}
\author{Frank Petriello}
\email[]{f-petriello@northwestern.edu}
\affiliation{High Energy Physics Division, Argonne National Laboratory, Argonne, IL 60439, USA} 
\affiliation{Department of Physics \& Astronomy, Northwestern University, Evanston, IL 60208, USA}
\author{Frank J. Tackmann}
\email[]{frank.tackmann@desy.de}
\affiliation{Theory Group, Deutsches Elektronen-Synchrotron (DESY), D-22607 Hamburg, Germany} 
\author{Jonathan R. Walsh}
\email[]{jwalsh@lbl.gov}
\affiliation{Ernest Orlando Lawrence Berkeley National Laboratory, University of California, Berkeley, CA 94720, U.S.A.}
\affiliation{Center for Theoretical Physics, University of California, Berkeley, CA 94720, U.S.A.}


  \vspace*{0.3cm}

\begin{abstract}
  \vspace{0.5cm}

Experimental analyses often use jet binning to distinguish between different kinematic regimes and separate contributions from background processes.  To accurately model theoretical uncertainties in these measurements, a consistent description of the jet bins is required.  We present a complete framework for the combination of resummed results for production processes in different exclusive jet bins, focusing on Higgs production in gluon fusion as an example.  We extend the resummation of the $H + 1$-jet cross section into the challenging low transverse momentum region, lowering the uncertainties considerably.  We provide combined predictions with resummation for cross sections in the $H + 0$-jet and $H + 1$-jet bins, and give an improved theory covariance matrix for use in experimental studies.  We estimate that the relevant theoretical uncertainties on the signal strength in the $H \to WW^*$ analysis are reduced by nearly a factor of 2 compared to the current value.

\end{abstract}

\maketitle

\section{Introduction}
\label{sec:intro}

The Higgs program at the LHC requires exquisite experimental measurements coupled with precise theoretical predictions.  These predictions include inclusive cross sections as well as cross sections with experimental selection cuts. One example is the $H\to WW^* \to \ell^+ \ell^- \nu \bar{\nu}$ analysis, where the events are binned by exclusive jet multiplicity~\cite{Aad:2012tfa,Aad:2012uub,Chatrchyan:2012ufa,Chatrchyan:2012ty}.  The power of the analysis comes from the 0-jet and 1-jet bins, where the $t\bar{t}$ background contamination is small and the jet binning allows for a reduction in the Standard Model (SM) $pp\to WW$ background.  The exclusive jet cross sections are selected with a transverse momentum veto on final state jets found with a jet algorithm such as anti-$\kt$~\cite{Cacciari:2008gp}, where jets with $p_T^{\rm jet} > \pTcut$ are vetoed.  Because the Higgs mass peak cannot be reconstructed, a prediction for the SM cross section that includes the experimental event selection cuts, including the jet veto, is used in the measurement of the Higgs signal strength.

Fixed-order predictions, next-to-next-to-leading order (NNLO) for the 0-jet bin and next-to-leading order (NLO) for the 1-jet bin, are currently used to determine the perturbative uncertainties in the expected signal cross section.  Predictions at fixed order in perturbation theory can suffer from large uncertainties when selection cuts are applied due in part to unresummed logarithms involving the relevant scales in the process: the veto scale $\pTcut$, the Higgs mass $m_H$, and the $p_T$ of the final state jet, $p_{TJ}$ (in the 1-jet bin).  By resumming these logarithms to all orders the perturbative uncertainties can be considerably reduced.

Whether fixed-order or resummed predictions are used for the cross sections in the 0-jet and 1-jet bins, the inputs to the experimental analysis require a consistent treatment of the cross sections and uncertainties in the different jet bins since their combination is used in the measurement.  Defining
\begin{align}
\sigma_\tot &: \text{ the inclusive cross section} \,, \\
\sigma_0 (\pTcut) &: \text{ the 0-jet cross section, with no jets with } \pTjet > \pTcut \,, \nn \\
\sigma_{\ge 1} (\pTcut) &: \text{ the inclusive 1-jet cross section, with at least 1 jet with } \pTjet > \pTcut \,, \nn \\
\sigma_1 ([p_{Ta}, p_{Tb}]; \pTcut) &: \text{ the exclusive 1-jet cross section, with } p_{Tb} > p_{TJ} > p_{Ta} \,, \nn \\
\sigma_{\ge 2} (\pTcut) &: \text{ the inclusive 2-jet cross section, with at least 2 jets with } \pTjet > \pTcut \,, \nn
\end{align}
simple consistency conditions relate these cross sections:
\begin{align}
\sigma_\tot &= \sigma_0 (\pTcut) + \sigma_{\ge 1} (\pTcut) \,, \nn \\
\sigma_{\ge 1} (\pTcut) &= \sigma_1 ([\pTcut, \infty]; \pTcut) + \sigma_{\ge 2} (\pTcut) \,.
\end{align}
These conditions imply that nontrivial correlations exist between the uncertainties in the different jet bins; for example, scale variation on the 0-jet cross section will feed through to the 1-jet cross section.  This is addressed by maintaining control of the correlations between the exclusive and inclusive jet bins in both the 0-jet and 1-jet predictions, allowing one to determine all of the entries in the covariance matrix $C(\{\sigma_0, \sigma_1, \sigma_{\ge 2}\})$, which can be used to calculate the perturbative theory uncertainties for any quantity built from these cross sections.

Two issues prevent the upgrade of the covariance matrix from a fixed-order one to a resummed one with reduced uncertainties.  The first issue is that the resummation of the 1-jet bin is only known when $p_{TJ} \gg \pTcut$, which leaves a large part of the 1-jet spectrum unresummed.  The second is that no method of consistently combining the resummations of the 0-jet and 1-jet bins that accounts for all correlations has been given. In this work we solve both of these outstanding problems. We first show how to improve the challenging low $p_{TJ}$ region of the 1-jet bin by resummation. We do so by using the inclusive 1-jet cross section, supplemented with a fixed-order correction from the 2-jet cross section, to provide a resummed prediction in the low $p_{TJ}$ regime.  We provide several cross-checks of this resummation, showing that it smoothly matches onto the resummed predictions at large $p_{TJ}$.  We then give a prescription for how to determine the covariance matrix $C(\{\sigma_0, \sigma_1, \sigma_{\ge 2}\})$ for both fixed-order and resummed predictions.  We numerically illustrate our results using the $H\to WW^* \to \ell^+ \ell^- \nu \bar{\nu}$ analysis at the LHC.  We estimate that the switch from a fixed-order covariance matrix to a resummed one reduces the theoretical uncertainty on the signal strength extracted by the experimental analyses by a factor of two.  This is a dramatic reduction, one which will significantly improve our ability to uncover the underlying nature of the Higgs boson.

Two other aspects of our results warrant emphasis.  First, the framework we introduce for the combination of the 0-jet and 1-jet bins is systematically improvable, and can incorporate improved perturbative information as it becomes available.  We demonstrate that point in this manuscript.  For example, since we are now able to resum the entire 1-jet spectrum and therefore include its dominant higher-order terms beyond NLO, we are able to use the NLO prediction for the inclusive 2-jet spectrum, which is currently treated at LO in experimental studies~\cite{Chatrchyan:2012ty}.  Second, although we focus on the $WW^*$ final state, the numerical improvement we find will hold for any Higgs analysis in which the inclusive 2-jet bin is analyzed separately from the other jet bins.  We also note that although we focus on Higgs production in this work, our framework is applicable to any other process in which the signal is divided according to jet multiplicity.

Our paper is organized as follows.  In \sec{jetbinuncertainties}, we discuss the general form of the theory covariance matrices for both fixed-order and resummed predictions.  In \sec{lowpTJ}, we show how the 0-jet resummed predictions along with fixed-order corrections may be used to make a resummed prediction for the low $p_{TJ}$ region of the exclusive 1-jet cross section.  In \sec{numerics}, we apply this framework to produce predictions for the cross section and uncertainties for the complete 0-jet and 1-jet bins.  We present an illustrative example that is phenomenologically relevant to the current $H \to WW^*$ analysis, and show how the uncertainties are reduced in the measurement of the signal strength.  Finally, in \sec{conclusions} we conclude.

\section{Perturbative Uncertainties in Jet Binning}
\label{sec:jetbinuncertainties}

Whether cross section predictions are fixed-order or resummed, scale variation is the primary tool for assessing theoretical uncertainties due to unknown higher-order perturbative corrections. In this section we discuss the general parametrization of perturbative uncertainties in exclusive jet bins in terms of the full theory covariance matrix. We then give the methods for how the uncertainties can be reliably estimated and how to determine the full covariance matrix for both fixed-order and resummed predictions.

\subsection{General Parametrization}
\label{subsec:general}

A convenient way to describe the perturbative uncertainties involved in the jet binning is in terms of fully correlated and fully anticorrelated components. Following Refs.~\cite{Stewart:2011cf, Gangal:2013nxa}, for a single jet bin boundary between $N$ and $N+1$ jets, this amounts to parametrizing the covariance matrix for $\{\sigma_N, \sigma_{\geq N+1}\}$ as
\begin{align} \label{eq:Cgeneral}
C(\{\sigma_N, \sigma_{\geq N+1}\}) &=
\begin{pmatrix}
(\Delta^{\rm y}_{N})^2 &  \Delta^{\rm y}_{N}\,\Delta^{\rm y}_{\geq N+1}  \\
\Delta^{\rm y}_{N}\,\Delta^{\rm y}_{\geq N+1} & (\Delta^{\rm y}_{\geq N+1})^2
\end{pmatrix}
+
\begin{pmatrix}
 \Delta_{N\,\cut}^2 &  - \Delta_{N\,\cut}^2 \\
-\Delta_{N\,\cut}^2 & \Delta_{N\,\cut}^2
\end{pmatrix}
,\end{align}
where the first contribution, denoted with a superscript ``y'', is interpreted as an overall yield uncertainty which by definition is 100\% correlated between the two bins $\sigma_N$ and $\sigma_{\geq N+1}$. The second contribution is the migration uncertainty which is anticorrelated between the bins. It corresponds to the uncertainty introduced by the binning which separates $N$ from $N+1$ jets, and by definition drops out in the sum $\sigma_N+\sigma_{\geq N+1}$.

For the combination of the exclusive 0-jet, exclusive 1-jet, and inclusive 2-jet bins, we choose to write the covariance matrix in the basis $\{\sigma_0, \sigma_1, \sigma_{\ge 2} \}$. We then have the general parametrization
\begin{equation} \label{eq:Cgen}
C(\{\sigma_0, \sigma_1, \sigma_{\ge 2}\}) =
C_{\rm y}(\{\sigma_0, \sigma_1, \sigma_{\ge 2}\})
+ C_\cut(\{\sigma_0, \sigma_1, \sigma_{\ge 2}\})
\,.\end{equation}
The yield uncertainties described by $C_{\rm y}$ are fully correlated between all jet bins,
\begin{equation} \label{eq:Cy}
C_{\rm y}(\{\sigma_0, \sigma_1, \sigma_{\ge 2}\}) =
\begin{pmatrix}
(\Delta_0^{\rm y})^2 \;&\; \Delta_0^{\rm y} \Delta_1^{\rm y} \;&\; \Delta_0^{\rm y} \Delta_{\ge2}^{\rm y} \\
\Delta_0^{\rm y} \Delta_1^{\rm y} \;&\; (\Delta_1^{\rm y})^2 \;&\; \Delta_1^{\rm y} \Delta_{\ge2}^{\rm y} \\
\Delta_0^{\rm y} \Delta_{\ge2}^{\rm y} \;&\; \Delta_1^{\rm y} \Delta_{\ge2}^{\rm y} \;&\; (\Delta_{\ge2}^{\rm y})^2
\end{pmatrix}
\,,\end{equation}
where $\Delta_0^{\rm y}$, $\Delta_1^{\rm y}$, and $\Delta_{\ge 2}^{\rm y}$ are the yield uncertainties for each jet bin.
The migration uncertainties are
\begin{equation} \label{eq:Ccut}
C_\cut(\{\sigma_0, \sigma_1, \sigma_{\ge 2}\})
= \begin{pmatrix}
\Delta_{0\,\cut}^2 \;&\; - \Delta_{0\,\cut}^2 + C_{01\,\cut} \;&\; - C_{01\,\cut} \\
- \Delta_{0\,\cut}^2 + C_{01\,\cut} \;&\; \Delta_{0\,\cut}^2 + \Delta_{1\,\cut}^2 - 2C_{01\,\cut} \;&\; - \Delta_{1\,\cut}^2 + C_{01\,\cut} \\
- C_{01\,\cut} \;&\; -\Delta_{1\,\cut}^2 + C_{01\,\cut}  \;&\; \Delta_{1\,\cut}^2
\end{pmatrix}
\,.\end{equation}
A priori, the migration uncertainties $\Delta_{0\,\cut}$ of the 0-jet boundary and $\Delta_{1\,\cut}$ of the 1-jet boundary can have a nontrivial correlation, which is encoded in the additional parameter $C_{01\,\cut}$.  The structure of $C_\cut$ is fixed by the requirement that in the sum $\sigma_1 + \sigma_{\geq 2}$ any dependence on the 1-jet boundary must drop out, while in the sum $\sigma_0 + \sigma_1$ any dependence on the 0-jet boundary must drop out. Together this automatically implies that in the total cross section any migration uncertainties drop out as they must, i.e., all elements of $C_\cut$ must sum to zero.

\subsection{Fixed-Order Predictions}
\label{subsec:FO}

At fixed order, a direct scale variation is typically used, where the renormalization and factorization scales are varied around central values. Using a common (correlated) scale variation for all jet bins amounts to setting $\Delta_{i\,\cut} = 0$ and $\Delta_i^{\rm y} = \Delta_i^\FO$. However, this can lead to artificial cancellations in exclusive cross sections in the regime where the logarithmic corrections are not small and the migration uncertainties cannot be neglected~\cite{Stewart:2011cf}.

One method to ameliorate this cancellation and obtain a more reliable uncertainty estimate in exclusive jet bins is to explicitly take into account an estimate of the migration uncertainty using the ansatz
\begin{equation}
\Delta_N^{\rm y} = \Delta^\FO_{\geq N}
\,,\qquad
\Delta_{\geq N+1}^{\rm y} = 0
\,,\qquad
\Delta_{N\,\cut} = \Delta^\FO_{\geq N+1}
\,,\end{equation}
where the uncertainties $\Delta^\FO_{\geq N+1}$ in the inclusive cross sections are determined by the usual direct scale variation. This so-called ``ST method'' was proposed in Ref.~\cite{Stewart:2011cf}, and has been adopted in various exclusive analyses at the LHC and Tevatron including the ATLAS and CMS Higgs analyses.  An alternative approach, the ``efficiency method'', was proposed in Ref.~\cite{Banfi:2012yh}.

This ansatz assumes the dominance of the large logarithmic corrections in the perturbative expansion. Since the yield and migration uncertainties are independent, the limit $\Delta_{N\,\cut} = \Delta^\FO_{\geq N+1}$ implies that the inclusive fixed-order uncertainties $\Delta^\FO_{\geq N}$ and $\Delta^\FO_{\geq N+1}$ are treated as uncorrelated. The inclusive $N$-jet and inclusive $M$-jet cross sections are sensitive to different logarithms.  The former depends on the cut separating the $N$-jet and ($N-1$)-jet bins, while the latter depends on the cut dividing the $M$-jet and ($M-1$)-jet bins, which are independent parameters for $M \neq N$.  As long as the logarithms are the dominant source of corrections, it is justified to treat these uncertainties as independent, which implies that $\Delta_{N\,\cut}$ and $\Delta_{M\,\cut}$ are assumed to be uncorrelated so $C_{NM\,\cut} = 0$.

In terms of the full covariance matrix $C_\FO$, the covariance matrix elements between two inclusive jet cross sections is then given by
\begin{align}
&\textit{Inclusive-Inclusive :} \nn \\
&\qquad (C_\FO)_{\ge i,\ge j}
= \begin{cases}
\; (\Delta_{\ge i}^\FO)^2 &\textrm{ if } j = i \,, \\
\; 0 &\textrm{ otherwise} \,,
\end{cases}
\end{align}
between inclusive and exclusive jet cross sections they are
\begin{align}
&\textit{Inclusive-Exclusive :} \nn \\
&\qquad (C_\FO)_{\ge i,j}
= \begin{cases}
\; (\Delta_{\ge i}^\FO)^2 &\textrm{ if } j = i \,, \\
\; -(\Delta_{\ge i}^\FO)^2 &\textrm { if } j = i-1 \,, \\
\; 0 &\textrm{ otherwise} \,,
\end{cases}
\end{align}
and between exclusive jet cross sections they are
\begin{align}
&\textit{Exclusive-Exclusive :} \nn \\
& \qquad (C_\FO)_{i,j}
= \begin{cases}
\; (\Delta_{\ge i}^\FO)^2 + (\Delta_{\ge i+1}^\FO)^2 &\textrm{ if } j = i \,, \\
\; -(\Delta_{\ge i}^\FO)^2 &\textrm { if } j = i-1 \,, \\
\; -(\Delta_{\ge i+1}^\FO)^2 &\textrm { if } j = i+1 \,, \\
\; 0 &\textrm{ otherwise} \,.
\end{cases}
\end{align}

For our case of interest, the resulting fixed-order covariance matrix in the basis $\{\sigma_0, \sigma_1, \sigma_{\ge 2} \}$ is
\be \label{eq:CFO}
C_\FO (\{\sigma_0, \sigma_1, \sigma_{\ge 2} \}) =
\begin{pmatrix}
(\Delta_{\ge 0}^\FO)^2 + (\Delta_{\ge 1}^\FO)^2 \;&\; -(\Delta_{\ge 1}^\FO)^2 \;&\; 0  \\
-(\Delta_{\ge 1}^\FO)^2 \;&\; (\Delta_{\ge 1}^\FO)^2 + (\Delta_{\ge 2}^\FO)^2 \;&\; -(\Delta_{\ge 2}^\FO)^2  \\
0 \;&\; -(\Delta_{\ge 2}^\FO)^2  \;&\; (\Delta_{\ge 2}^\FO)^2
\end{pmatrix}.
\ee
Here, each of the $\Delta_{\geq i}^\FO$ is determined by direct scale variation in the corresponding inclusive fixed-order cross section $\sigma_{\geq i}$.  The inclusive 1-jet and 2-jet cross sections, $\sigma_{\geq 1}$ and $\sigma_{\geq 2}$, depend on the choice of $\pTcut$ as well as the jet algorithm.

\subsection{Resummed Predictions}
\label{subsec:resum}

Resummed predictions have more handles to estimate the perturbative uncertainties through scale variation of the various factorization and matching scales present in the resummed cross section. This means that one generally has finer control over the uncertainties, but also that care must be taken not to overestimate or underestimate them.

In a given resummed jet bin, one can identify two different types of uncertainties, which are defined precisely in Ref.~\cite{Stewart:2013faa}: First, an overall fixed-order component of the uncertainty, $\Delta_i^\mu$, which is estimated by collectively varying all of the scales that appear in the resummed prediction.  This variation is therefore insensitive to the logarithms of the various scale ratios in the result. Instead, it probes the nonlogarithmic contributions and reduces to the usual fixed-order scale variation in the limit of large $\pTcut$. This component is identified with the yield uncertainty, $\Delta_i^{\rm y} = \Delta_i^\mu$, so
\begin{equation} \label{eq:Cmu}
C_{\rm y}(\{\sigma_0, \sigma_1, \sigma_{\ge 2}\}) \equiv 
C_{\mu} (\{\sigma_0, \sigma_1, \sigma_{\ge 2}\}) =
\begin{pmatrix}
(\Delta_0^\mu)^2 \;&\; \Delta_0^\mu \Delta_1^\mu \;&\; \Delta_0^\mu \Delta_{\ge2}^\mu \\
\Delta_0^\mu \Delta_1^\mu \;&\; (\Delta_1^\mu)^2 \;&\; \Delta_1^\mu \Delta_{\ge2}^\mu \\
\Delta_0^\mu \Delta_{\ge2}^\mu \;&\; \Delta_1^\mu \Delta_{\ge2}^\mu \;&\; (\Delta_{\ge2}^\mu)^2
\end{pmatrix},
\end{equation}
Since the cross sections in these jet bins sum to the total cross section, the yield uncertainties sum to the uncertainty in the total cross section, $\Delta_\tot^\mu$:
\begin{equation}
\Delta_\tot^\mu = \Delta_0^\mu + \Delta_1^\mu + \Delta_{\ge 2}^\mu \,.
\end{equation}
For our purposes this relation defines $\Delta_{\ge 2}^\mu$.

Second, the resummed component of the uncertainty, $\Delta_i^\res$, is estimated by varying the individual resummation scales in separate directions and probes the impact of the logarithmic corrections in the cross section.  Since the migration between jet bins is sensitive to the logarithm of the corresponding jet-bin separation parameter (i.e., the $\pTcut$ value), this motivates the identification of the resummed component of the scale variation with the anti-correlated migration uncertainty,
\begin{equation} \label{eq:Cres}
C_\cut(\{\sigma_0, \sigma_1, \sigma_{\ge 2}\}) \equiv  C_\res(\{\sigma_0, \sigma_1, \sigma_{\ge 2}\})
\,.\end{equation}
The precise determination of the elements $\Delta_{0\,\cut}$, $\Delta_{1\,\cut}$, and $C_{01\,\cut}$ in our case is nontrivial and discussed further in \sec{lowpTJ}.

\subsection{Current Status of Predictions}
\label{subsec:status}

The full covariance matrix in \eq{Cgen}, determined either in fixed order as in \eq{CFO} or via resummation, allows one to simultaneously describe the theoretical uncertainties associated with the perturbative expansion for the 0-jet and 1-jet bins.  For Higgs production, we briefly summarize the results which are available to determine the elements of these matrices.

At fixed order, the $H+0$-jet cross section is known at NNLO~\cite{Anastasiou:2005qj,Catani:2007vq} and the $H+1$-jet cross section is known at NLO~\cite{Glosser:2002gm,Ravindran:2002dc}.  The $H+1$-jet cross section at NNLO is currently being computed using new techniques~\cite{Czakon:2010td,Boughezal:2011jf}, with results available in the $gg\to Hg$ channel~\cite{Boughezal:2013uia}.  The full NNLO result can be expected to substantially lower the uncertainty of the NLO result.  All of these results are in the effective theory obtained by integrating out the top quark.  Several theoretical refinements beyond this approximation are also available for these cross sections, including electroweak effects~\cite{Aglietti:2004nj,Actis:2008ug,Anastasiou:2008tj,Keung:2009bs,Brein:2010xj,Passarino:2013nka} and finite quark-mass corrections~\cite{Spira:1995rr,Harlander:2009my,Harlander:2012hf}.  In our results we rescale the 0-jet and 1-jet cross sections by the ratio of the leading-order $H+0$-jet cross section between the full and the effective theories, which is known to give a good approximation to the quark-mass effects.  For simplicity we do not include electroweak corrections; they are small, at the percent-level, and do not modify any conclusion of our study.

Predictions for the resummed 0-jet cross section are available from several groups~\cite{Banfi:2012yh,Banfi:2012jm,Becher:2012qa,Becher:2013xia,Tackmann:2012bt,Stewart:2013faa,Banfi:2013eda}.  Each includes resummation at least through NNLL and fixed-order matching to NNLO, and the results are generally in good agreement.  For the 1-jet cross section an additional scale $p_{TJ}$ is present, complicating the factorization theorem.  In the high $p_T$ regime, $p_{TJ} \sim m_H$, the factorization theorem is known, resummation can be performed, and results for the 1-jet cross section are available through the $\text{NLL}^{\prime}+\text{NLO}$ level~\cite{Liu:2012sz,Liu:2013hba}.   We note that non-global logarithms~\cite{Dasgupta:2001sh} first occur in the 1-jet cross section at the $\text{NLL}^{\prime}$ level.  Their numerical impact was estimated in Ref.~\cite{Liu:2013hba} to be at or below the percent level of the exclusive $H+1$-jet cross section for the relevant values of $m_H$ and $\pTcut$, and therefore do not affect phenomenology.

In the small $p_{TJ}$ regime ($\pTcut \lesssim p_{TJ} \ll m_H$), a complete factorization theorem has not yet been derived and only fixed-order results are available so far.  This complicates predictions for the combined 0-jet and 1-jet bins.  The 1-jet cross section is largest in the low $p_{TJ}$ regime, and the fixed-order prediction in that regime has large uncertainties.  It is also challenging to combine the fixed order with the resummed predictions across the remainder of phase space, as a hybrid between \eq{CFO} and \eqs{Cmu}{Cres} would be needed to describe the uncertainties and correlations between jet bins.  In the next section we will study this transition in more detail, using the 0-jet results to provide a resummed prediction in the low $p_{TJ}$ regime.

\section{The Transition Between 0-jet and 1-jet Bins}
\label{sec:lowpTJ}

\subsection{Construction of the 1-jet Bin}
\label{subsec:1jetbin}

As discussed in the previous section, the direct resummation of the exclusive 1-jet bin can only be performed when $p_{TJ}$ is much larger than $\pTcut$ and of order $m_H$.  We therefore divide the low $p_{TJ}$ and high $p_{TJ}$ regimes using a parameter $\pToff > \pTcut$:
\be
\sigma_1 ([\pTcut, \infty]; \pTcut) = \sigma_1 ([\pTcut, \pToff]; \pTcut) + \sigma_1 ([\pToff, \infty]; \pTcut) \,.
\ee
If an event has $p_{TJ} < \pTcut$ then it is in the 0-jet bin.  In practice, $\pToff$ is taken to be around $m_H/2$.  The second term in this relation can be directly resummed to NLL$'$+NLO using the results of Refs.~\cite{Liu:2012sz, Liu:2013hba}.
To obtain a result for the first term that is improved beyond fixed order by resummation, we note that the following identity can be used to relate the exclusive 1-jet cross section to the inclusive 1-jet and inclusive 2-jet cross sections:
\be \label{eq:sigma1lowpTJ}
\sigma_1 ([\pTcut, \pToff]; \pTcut) = \bigl[ \sigma_{\ge 1} (\pTcut) - \sigma_{\ge 1} (\pToff) \bigr] - \bigl[ \sigma_{\ge 2} (\pTcut, \pTcut) - \sigma_{\ge 2} (\pToff, \pTcut) \bigr] \,.
\ee
The first bracketed set of terms give the cross section for one or more jets between $\pTcut$ and $\pToff$.  The second bracketed set of terms provides a correction to have only one jet in this $p_T$ range. Using the relation in \eq{sigma1lowpTJ}, we can construct a (partially) resummed prediction for $\sigma_1 ([\pTcut, \pToff]; \pTcut)$, as the inclusive 1-jet cross section in the first brackets can be obtained from the resummed 0-jet cross section, which is known to high accuracy~\cite{Banfi:2012yh,Banfi:2012jm,Becher:2012qa,Becher:2013xia,Tackmann:2012bt,Stewart:2013faa,Banfi:2013eda}.
The inclusive 2-jet cross sections in the second bracket can be calculated at fixed order up to NLO.

The difference of inclusive 1-jet cross sections is the same as the following difference of 0-jet cross sections,
\be
\sigma_{\ge 1} (\pTcut) - \sigma_{\ge 1} (\pToff) = \sigma_0 (\pToff) - \sigma_0 (\pTcut) \,.
\ee
For the resummation of the 0-jet terms, we use the NNLL$'$+NNLO results of Ref.~\cite{Stewart:2013faa}. Their difference describes the inclusive 1-jet rate through NLO plus an all-orders series of the inclusive 1-jet logarithms of $\pTcut / m_H$ and $\pToff / m_H$.  The 2-jet terms are calculated at NLO. This means that the leading missing terms in the exclusive 1-jet cross section appear at NNLO and come from the inclusive 1-jet contribution.  These corrections can contain at most a single logarithm of the ratios $\pTcut / m_H$ and $\pToff / m_H$ as well as possibly large nonlogarithmic corrections.  However, we will show that the resummation of the 0-jet terms can capture NNLO 1-jet terms that are known to be large, and our predictions may be tested against the complete $H + 1$-jet NNLO cross section once the calculation is complete~\cite{Boughezal:2013uia}.  There are additional unresummed corrections in the logarithmic series of the exclusive 1-jet cross section induced by the cut on the second jet. These are higher-order terms that convert the resummation from the inclusive to the exclusive 1-jet cross section (or equivalently correspond to a resummation of the inclusive 2-jet cross sections in \eq{sigma1lowpTJ}).  These terms enter at N$^3$LO and beyond in the exclusive 1-jet cross section, and we will provide evidence that their contribution is small.

We therefore have two approaches to obtain the exclusive 1-jet rate, $\sigma_1 ([\pTcut, \infty]; \pTcut)$:
\begin{itemize}
\item Direct evaluation of the exclusive 1-jet cross section.  The resummation can be reliably performed for $p_{TJ} \sim m_H$, and is matched onto the fixed-order cross section at $p_{TJ} = \pToff$.  The NLL$'$+NLO resummation in Ref.~\cite{Liu:2013hba} will be used for this direct approach.
\item Indirect evaluation of the exclusive 1-jet cross section for $p_{TJ} < \pToff$.  \eq{sigma1lowpTJ} is used in this case, with the inclusive 1-jet (equivalently, the 0-jet) terms resummed to NNLL$'$+NNLO using the results of Ref.~\cite{Stewart:2013faa}.  The 2-jet terms are calculated at NLO.
\end{itemize}
The combination of the indirect approach for $p_{TJ} < \pToff$ and the direct approach for $p_{TJ} > \pToff$ will allow for a description of the complete 1-jet bin, i.e.,
\be \label{eq:1jetmaster}
\sigma_1 ([\pTcut, \infty]; \pTcut) = \sigma_1^{\rm indirect} ([\pTcut, \pToff]; \pTcut) + \sigma_1^{\rm direct} ([\pToff, \infty]; \pTcut) \,.
\ee
Note that the uncertainty in this combination has to take into account the nontrivial correlations between the 0-jet and 1-jet resummation uncertainties due to the terms that make up the indirect contribution to the 1-jet rate.

\subsection{Uncertainties}
\label{subsec:unc}

First, we must address how to treat the scale variation of the fixed-order 2-jet components of \eq{sigma1lowpTJ}.  Should their scale variations be assigned to the fixed-order uncertainty or the resummation uncertainty, and how are their scale variations correlated with those of the resummed predictions?  Here we follow the same arguments as in the original ST method and assign them to the migration uncertainty.  To determine their correlation with the other pieces, we can consider the limit $\pToff \to \infty$, which corresponds to resumming none of the exclusive 1-jet bin.  We demand that we reproduce the result of the ST method for pure fixed-order uncertainties in this limit.  This leads to the conclusion that the scale variations of the 2-jet cross sections are uncorrelated with the 0-jet prediction.  The correlations between the 2-jet components of \eq{sigma1lowpTJ} and the resummed 1-jet spectrum in the high $p_{Tj}$ region is not explicitly determined by these limits.  For simplicity, we set this correlation to zero.

The yield uncertainties entering $C_{\rm y}$ in \eq{Cy} are then given by
\begin{align} \label{eq:Delta1mu}
\Delta_0^{\rm y} &\equiv \Delta_0^\mu(\pTcut)
\,,\nn\\
\Delta_1^{\rm y} &\equiv \Delta_1^\mu([\pTcut, \infty]; \pTcut) = \Delta_0^\mu (\pToff) - \Delta_0^\mu (\pTcut) + \Delta_1^\mu ([\pToff, \infty]; \pTcut)
\,,\nn\\
\Delta_{\geq 2}^{\rm y} &\equiv \Delta_\tot^\mu - \Delta_0^\mu(\pTcut) - \Delta_1^\mu([\pTcut, \infty]; \pTcut)
\,.\end{align}
Since the yield uncertainties are fully correlated, the yield uncertainty in the 1-jet bin is the linear sum from the two regions, and for the region below $\pToff$ we used $\Delta_1^\mu ([\pTcut, \pToff]; \pTcut) = \Delta_0^\mu (\pToff) - \Delta_0^\mu (\pTcut)$. The individual contributions are evaluated as follows:
\begin{align}
\Delta_0^\mu(\pTcut),\, \Delta_0^\mu(\pToff),\, \Delta_\tot^\mu &: \text{ determined by overall scale variation in Ref.~\cite{Stewart:2013faa}.}
\nn \\
\Delta_1^\mu ([\pToff, \infty]; \pTcut) \; &: \text{ determined by overall scale variation in Ref.~\cite{Liu:2013hba}.}
\end{align}

Next, the migration uncertainties entering $C_\cut$ in \eq{Ccut} are given by
\begin{align} \label{eq:Delta1res}
\Delta_{0\,\cut} &\equiv \Delta_0^\res(\pTcut)
\,,\nn\\
\Delta_{1\,\cut}^2 &\equiv [\Delta_0^\res(\pToff)]^2 + \bigl[\Delta^\FO_{\ge 2} (\pToff, \pTcut) \bigr]^2 + \bigl\{ \Delta_1^\res ([\pToff, \infty]; \pTcut) \bigr\}^2
\,,\nn\\
C_{01\,\cut} &\equiv \Delta_0^\res (\pTcut) \Delta_0^\res (\pToff)
\,,\end{align}
where the individual contributions are evaluated as follows:
\begin{align}
\Delta_0^\res (\pTcut),\, \Delta_0^\res (\pToff) &\; : \text{ determined by resummation scale variations in Ref.~\cite{Stewart:2013faa}.}
\nn \\
\Delta_1^\res ([\pToff, \infty]; \pTcut) &\; : \text{ determined by resummation scale variations in Ref.~\cite{Liu:2013hba}.}
\nn \\
\Delta^\FO_{\ge 2} (\pToff, \pTcut) &\; : \text{ determined by fixed-order scale variation.}
\end{align}
The inclusive 2-jet uncertainty $\Delta^\FO_{\ge 2}$ is given by scale variation of the 2-jet fixed-order contributions in \eq{sigma1lowpTJ}.

The first two terms in $\Delta_{1\,\cut}$ corresponds to the $1$-jet migration uncertainty for the $p_{TJ} < \pToff$ region. Note that $\Delta_0^\res(\pToff)$ contributes to $\Delta_{1\,\cut}$ because it removes cross section from the fully inclusive $\sigma_{\geq 1}(\pTcut)$ making it more exclusive. The third term in $\Delta_{1\,\cut}$ is the $1$-jet migration uncertainty for the $p_{TJ} > \pToff$ region. The migration uncertainties between the two regions of $p_{TJ}$ are uncorrelated.
Finally, the result for $C_{01\,\cut}$ in \eq{Delta1res}, encoding the correlation between $\Delta_{0\,\cut}$ and $\Delta_{1\,\cut}$, is determined as follows: the high $p_{TJ}$ region of the 1-jet bin is directly resummed, and is therefore uncorrelated with the 0-jet bin.  The low $p_{TJ}$ region of the 1-jet bin contains a contribution from the difference of 0-jet cross sections, as shown in \eq{sigma1lowpTJ}, and $\Delta_0^\res (\pTcut)$ and $\Delta_0^\res(\pToff)$ should be treated as correlated, which yields $C_{01\,\cut}$. To see this explicitly, only considering the $p_{TJ} < \pToff$ contributions, the total 1-jet migration uncertainty is
\begin{equation}
[\pTcut ,\, \pToff] \,:\; (C_\cut)_{1,1}
= \bigl[ \Delta_0^\res (\pToff) - \Delta_0^\res (\pTcut) \bigr]^2  + \bigl[\Delta^\FO_{\ge 2} (\pToff, \pTcut) \bigr]^2
\equiv \Delta_{0\,\cut}^2 + \Delta_{1\,\cut}^2 - 2C_{01\,\cut}
\,.\end{equation}

Next, we will study the indirect approach to the cross section in the 1-jet bin for low $p_{TJ}$ to show that it provides a valid description in that regime.

\subsection{Testing the Indirect Approach}
\label{subsec:indirect}

To show that \eq{sigma1lowpTJ} improves upon the fixed-order description of the low $p_{TJ}$ region, we must demonstrate that a few criteria are fulfilled.  First, the contributions from the inclusive 2-jet cross section must be small compared to the 0-jet terms, so that the resummation can be performed for the bulk of the contribution.  Since we will switch to the direct resummation of the 1-jet cross section above $\pToff$, we must also show that the two predictions smoothly match onto each other at that point.  Finally, we must demonstrate that these first two points are insensitive to the exact numerical value chosen for $\pToff$.

\begin{figure}[!t]
\begin{center}
\includegraphics[width=0.65\textwidth,angle=90]{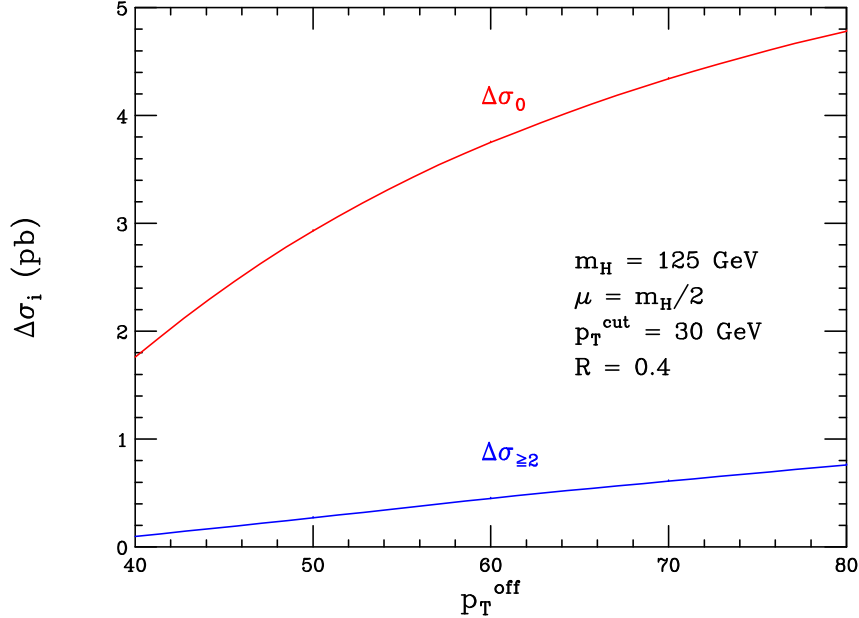}
\end{center}
\vspace{-1.0cm}
\caption{The fixed-order cross sections $\Delta \sigma_0$ and $\Delta \sigma_{\geq 2}$ defined in Eq.~(\ref{eq:sigma1lowpTJcomp}) as a function of $\pToff$, with all other parameters as given in the figure. } \label{fig:poff}
\end{figure}

To address the first point, we compute the relevant cross sections at fixed order using MCFM~\cite{Campbell:2010ff, Campbell:2010cz}.  We use the following set of parameters:
\begin{align}
m_H &= 125 \GeV \,, \qquad \pTcut = 30 \GeV \,, \nn \\
\mu_R = \mu_F \equiv \mu &= m_H / 2 \,, \qquad\qquad\! R = 0.4 \,.
\end{align}
The choice of $m_H/2$ as the central scale choice has been previously argued to be the appropriate scale choice when evaluating Higgs cross sections at fixed order~\cite{Dittmaier:2011ti}.  We use MSTW NNLO PDFs~\cite{Martin:2009iq}, and set the LHC collision energy to 8 TeV.  Defining
\bea \label{eq:sigma1lowpTJcomp}
\Delta \sigma_0 &=& \sigma_0 (\pToff) - \sigma_0 (\pTcut) = \sigma_{\ge 1} (\pTcut) - \sigma_{\ge 1} (\pToff) \,, \nonumber \\
\Delta \sigma_{\geq 2} &=& \sigma_{\ge 2} (\pTcut, \pTcut) - \sigma_{\ge 2} (\pToff, \pTcut) \,,
\eea
we plot in \fig{poff} these two contributions to the low-$p_{TJ}$ 1-jet region as a function of $\pToff$ in the region $\pToff \sim m_H/2$.  We have evaluated $\Delta \sigma_0$ at NLO and $\Delta \sigma_{\geq 2}$ at LO in the fixed-order perturbative expansion, so that both contain terms through $\ord{\as^2}$.  The inclusive 2-jet correction remains a small fraction of $\Delta \sigma_0$, varying from 6\% at $\pToff=40$ GeV to 16\% at $\pToff=80$ GeV.  This implies that the higher-order terms in the 2-jet cross section, those required to convert the resummation of the inclusive 1-jet logarithms to the exclusive rate, are small.  We therefore conclude that resumming only the 0-jet component of \eq{sigma1lowpTJ} may constitute an improvement of the low $p_{TJ}$ prediction.

We next study whether the improvement of the low $p_{TJ}$ region proposed in \eq{sigma1lowpTJ} smoothly matches onto the direct resummation of the high $p_{TJ}$ region near $\pToff$.  For the 0-jet component of \eq{sigma1lowpTJ} we use the $\text{NNLL}^{\prime}+\text{NNLO}$ results of Ref.~\cite{Stewart:2013faa}, while for the 1-jet prediction we use the $\text{NLL}^{\prime}+\text{NLO}$ results of Ref.~\cite{Liu:2013hba}.  The choices for the central values of the various scales which appear in each resummed calculation, together with the procedure for varying these scales around their central values to obtain uncertainties, are described in detail in the original papers.  We use the NLO prediction for the inclusive 2-jet piece of \eq{sigma1lowpTJ}, now with $\mu_R = \mu_F = m_H$ since we are resumming the bulk of the cross section.

One feature we wish to additionally demonstrate with this comparison is that the proposal of \eq{sigma1lowpTJ} represents a systematically-improvable framework that can incorporate future theoretical advances as they become available.  We therefore include the 0-jet component of \eq{sigma1lowpTJ} both with and without complex scale setting, which
is known to resum large $\pi^2$ terms in the $gg\to H$ form factor relevant for Higgs production~\cite{Parisi:1979xd, Sterman:1986aj, Magnea:1990zb, Ahrens:2008qu, Ahrens:2008nc}.  Similarly, the NNLO hard function (see Ref.~\cite{Gehrmann:2011aa}) is known to induce a large correction to the gluon-fusion channel of the $H+1$-jet cross section~\cite{Boughezal:2013uia}.  From these results we can find that for the scale choice $\mu_R = \mu_F = m_H$, the hard function gives a correction to the NLO 1-jet cross section of approximately 30\%, compared to a total correction of 40\% when going from NLO to NNLO.  It is the dominant source of the NNLO correction for these parameter choices.  Furthermore, the enhancement from the hard function is constant over a large range of $p_{TJ}$, suggesting that its origin is a large constant term similar to the $\pi^2$ corrections that the complex scale setting resums for the 0-jet cross section.  We therefore extend the resummation of  Ref.~\cite{Liu:2013hba} to include the NNLO $H+1$-jet hard function.   We compare the matching of the low $p_{TJ}$ and high $p_{TJ}$ regions in two schemes:
\begin{itemize}
\item Scheme A: complex scale setting for the 0-jet terms in the indirect approach, and the direct 1-jet cross section with the NNLO hard function incorporated.
\item Scheme B: no complex scale setting for the 0-jet terms in the indirect approach, and the direct 1-jet cross section with only the NLO hard function.
\end{itemize}
Scheme A will lead to larger cross sections due to the $\pi^2$ corrections and the NNLO hard function, whose effects on the indirect and direct cross sections are roughly equivalent.  Scheme A incorporates more known information than Scheme B, and is therefore our preferred choice for matching the 0-jet and 1-jet bins.  We will show that it leads to a smoother matching, and that it is also more independent of the parameter $\pToff$ that is used to separate the two regions.

\begin{figure}
\centering
\mbox{\subfigure{\includegraphics[width=3.1in]{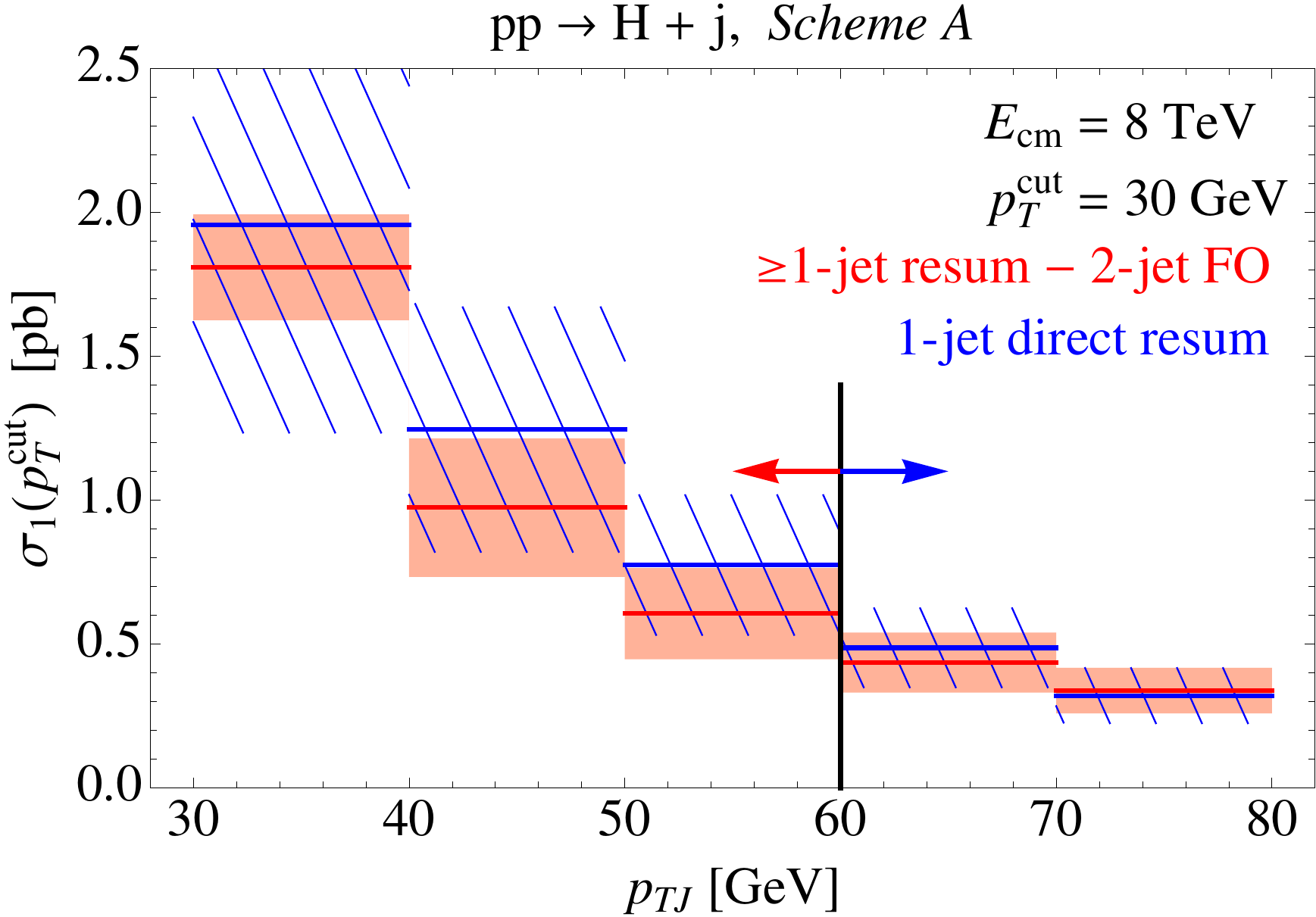}}\quad
\subfigure{\includegraphics[width=3.1in]{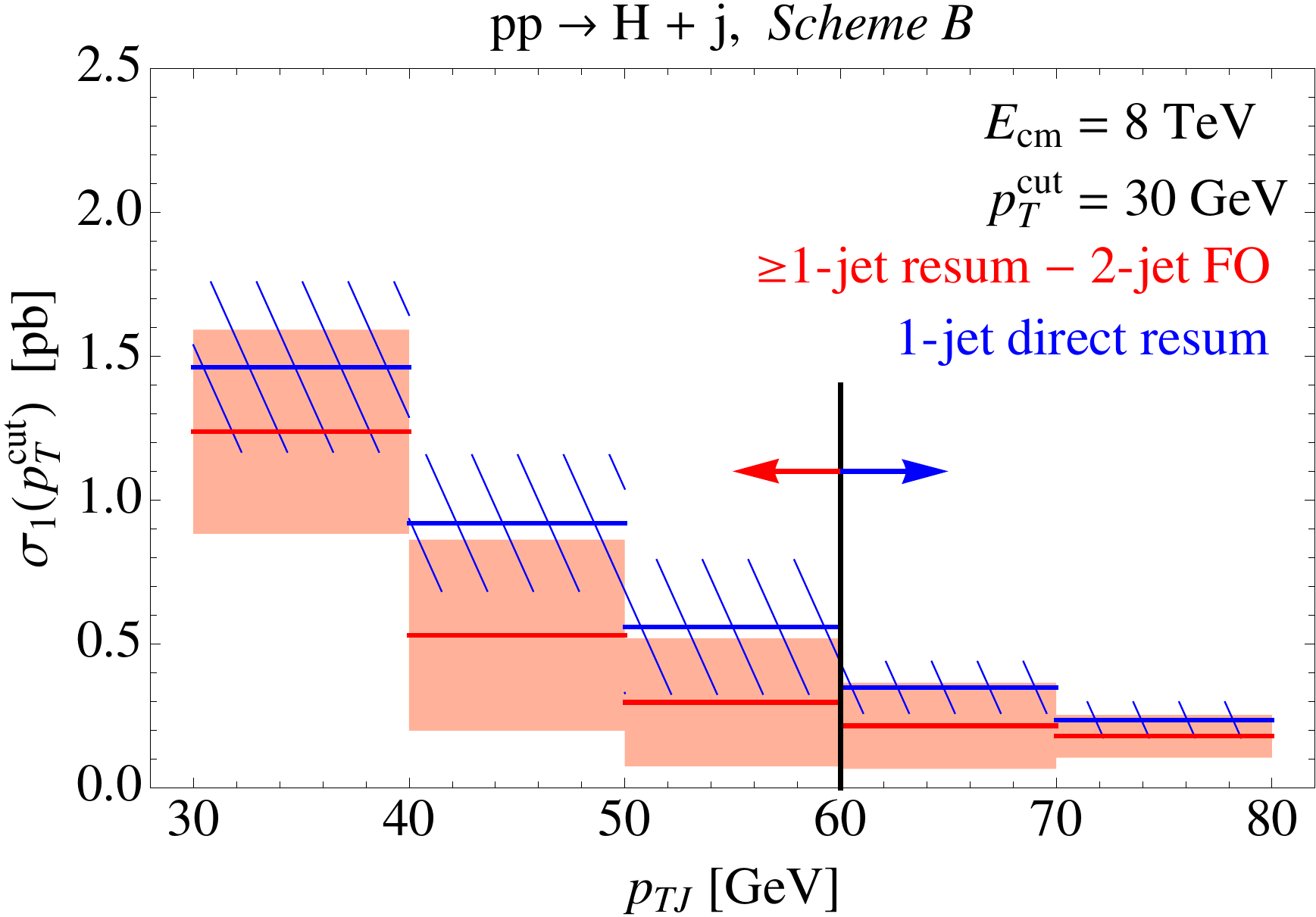} }}
\caption{Comparison of the 1-jet cross section in bins of $p_{TJ}$.  The direct (blue, hatched error bands) and indirect (red, solid error bands) approaches are shown.  Scheme A is on the left, B is on the right (the two schemes are described in the text).  The line at $\pToff = 60 \GeV$ indicates that the indirect approach is used for $p_{TJ} < \pToff$ and the direct approach is used for $p_{TJ} > \pToff$.} \label{fig:comparison}
\end{figure}

To make predictions in bins of $p_{TJ}$, we must extend \eq{sigma1lowpTJ}, which is only valid for a bin whose lower boundary is the veto scale $\pTcut$.  The correct generalization comes from the difference of \eq{sigma1lowpTJ} with different values of $\pToff$:
\begin{align}
\sigma_1 ([p_{Ta}, p_{Tb}]; \pTcut) &= \bigl[ \sigma_{\ge 1} (p_{Ta}) - \sigma_{\ge 1} (p_{Tb}) \bigr] - \bigl[ \sigma_{\ge 2} (p_{Ta}, \pTcut) - \sigma_{\ge 2} (p_{Tb}, \pTcut) \bigr] \,.
\end{align}
We compare the direct and indirect cross sections in bins of $p_{TJ}$ of 10 GeV width between 30 GeV and 80 GeV.  The results are shown in \fig{comparison}.  
In both schemes, the indirect and direct 1-jet cross sections are in relatively good agreement.  The large uncertainties at low $p_{TJ}$ in the direct evaluation of the 1-jet cross section are reduced by resumming the inclusive 1-jet logarithms in the indirect prediction.  While these approaches agree within uncertainties for Scheme B, there is a sizable offset between them.  Also, the uncertainties without complex scale setting in the 0-jet result are large.  The matching is significantly improved in Scheme A; the central values are closer, and the uncertainties of the indirect approach are smaller.  We note that the scale variation of the direct approach increases below $\pToff$ due to large logarithms $\text{ln}(m_H/p_{TJ})$ appearing in the NNLO hard function at low $p_{TJ}$.  The smooth matching between these predictions suggests that the indirect evaluation of the 1-jet cross section is valid.  We will use it for $p_{TJ} < \pToff$ and the direct approach for $p_{TJ} > \pToff$, as in \eq{1jetmaster}.

\begin{figure}
\centering
\mbox{\subfigure{\includegraphics[width=3.1in]{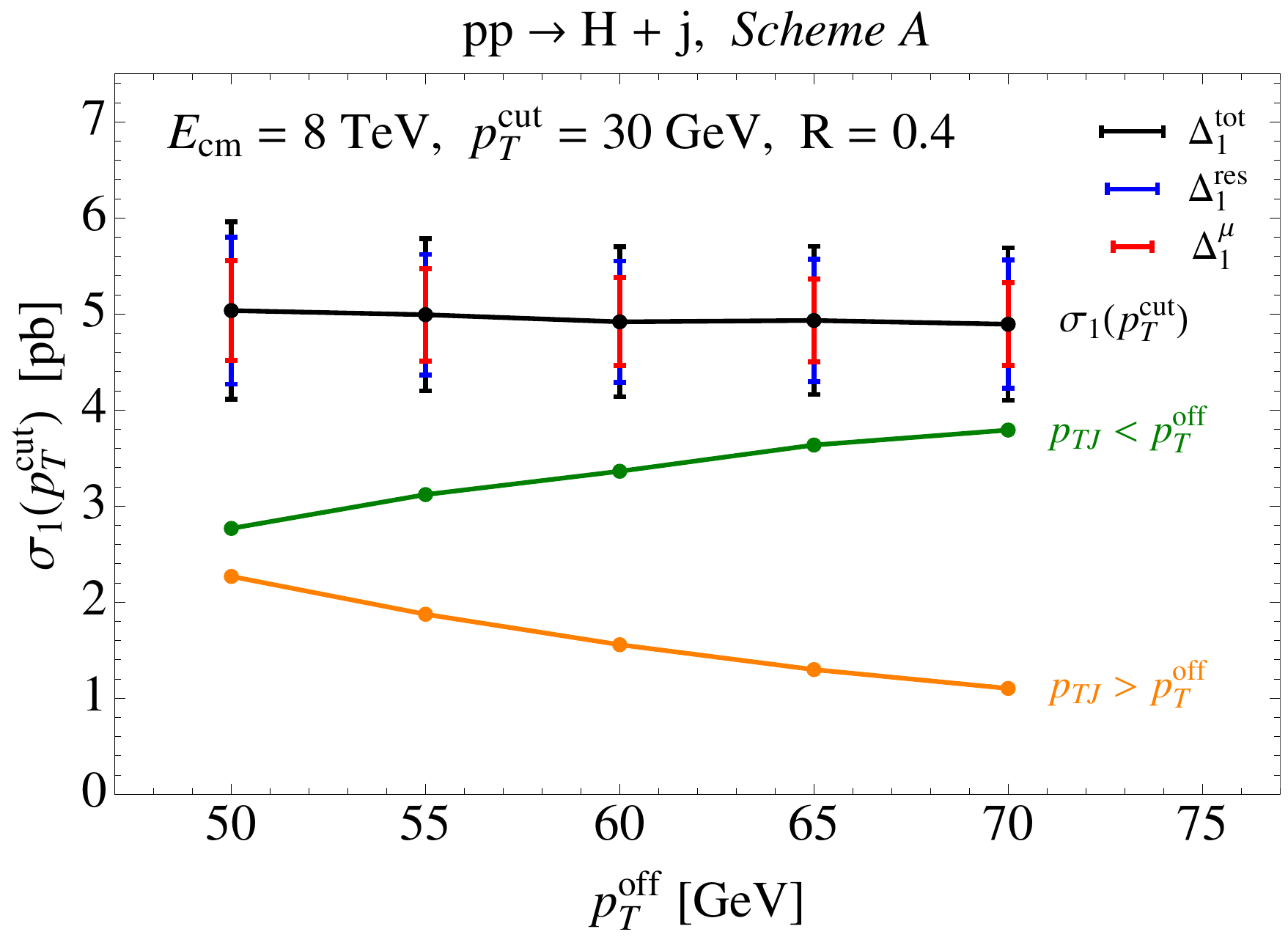}}\quad
\subfigure{\includegraphics[width=3.1in]{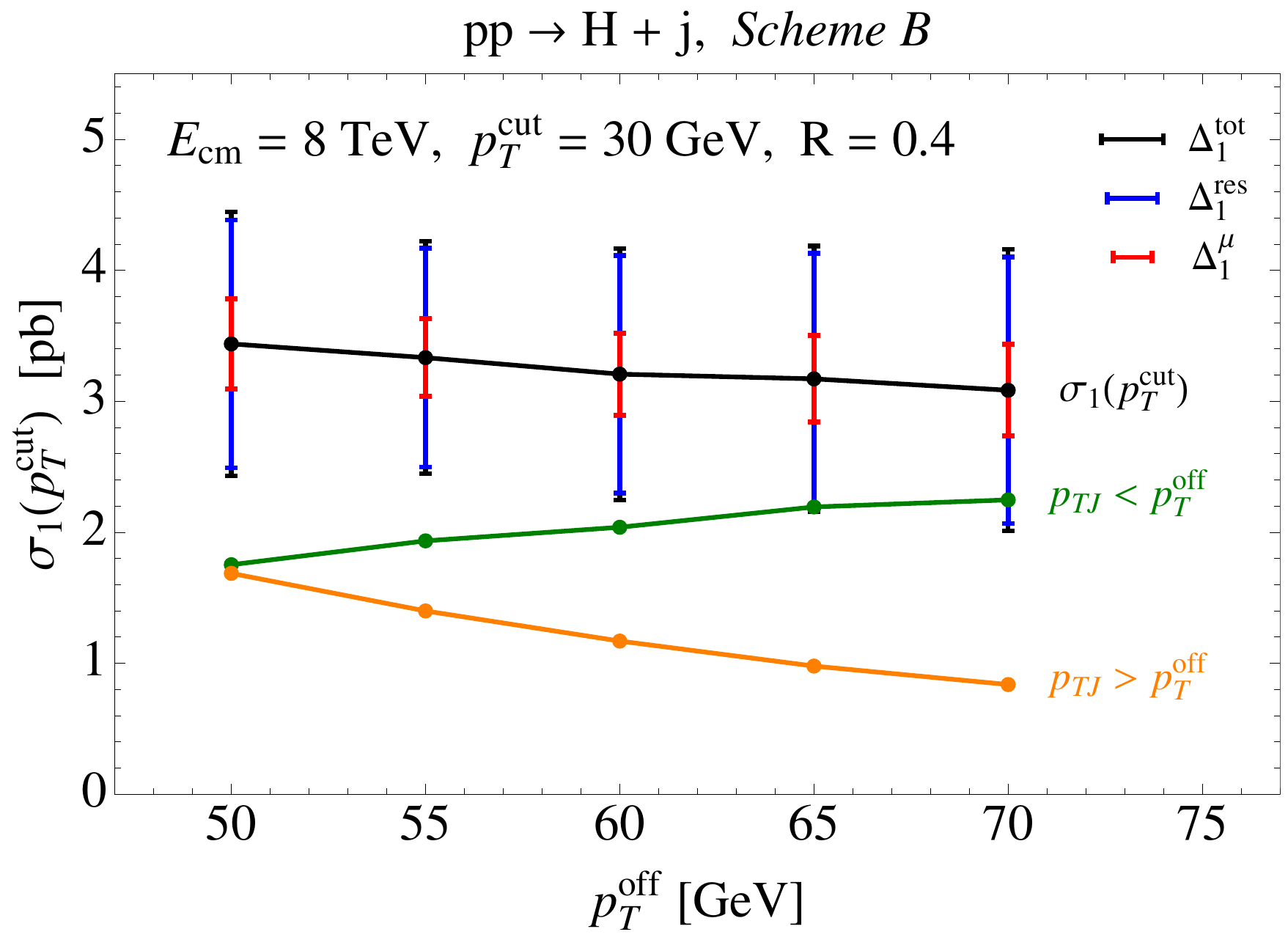} }}
\caption{Dependence on the parameter $\pToff$, which is used to divide the indirect (used for $p_{TJ} < \pToff$) and direct (used for $p_{TJ} > \pToff$) evaluations of the 1-jet cross section.  The black points give the total 1-jet rate (with total, yield, and migration uncertainties shown), and the green and orange points show the indirect and direct contributions respectively.} \label{fig:pToffdep}
\end{figure}
Finally, we show the $\pToff$ dependence of the 1-jet bin in \fig{pToffdep}.  The 1-jet cross section and its uncertainties are very insensitive to $\pToff$, with the net dependence much smaller than the change in the direct and indirect contributions to the 1-jet cross sections over the range of $\pToff$ studied.  This gives good confidence that the 1-jet bin is being accurately described by our method.  We note that the dependence on $\pToff$ is significantly reduced in Scheme A, providing further motivation for the choice of this scheme as the preferred way to match the 0-jet and 1-jet predictions.

In the next section we will perform a detailed numerical study of the 0-jet, 1-jet, and inclusive 2-jet cross sections, their uncertainties, and the covariance matrices for these predictions.

\section{Numerical Results for $H + 0, 1$ jets}
\label{sec:numerics}

We now have a consistent framework which incorporates resummation-improved predictions, and accounts for the correlations between their uncertainties, for the 0-jet and 1-jet bins of Higgs production.  We summarize the ingredients below.
\begin{itemize}

\item In the 0-jet bin, we use the direct resummation of the cross section as implemented in Ref.~\cite{Stewart:2013faa}, currently known to $\text{NNLL}^{\prime}+\text{NNLO}$.  We include the effect of complex scale setting in this prediction.

\item In the high $p_{TJ}$ region of the 1-jet bin, we use the direct resummation of Ref.~\cite{Liu:2013hba}, currently known to $\text{NLL}^{\prime}+\text{NLO}$.  We include the NNLO hard function in this prediction.

\item In the low $p_{TJ}$ region of the 1-jet bin, we use Eq.~(\ref{eq:sigma1lowpTJ}).  The 0-jet components of this equation are obtained from the 0-jet resummation, while the inclusive 2-jet pieces are obtained from MCFM at NLO.  We note that since we are now able to resum the entire 1-jet cross section and consequently include its dominant corrections beyond NLO, it is consistent to use the NLO result for the inclusive 2-jet piece rather than the LO one.

\end{itemize} 
For the covariance matrix, we can directly use the results of Eqs.~(\ref{eq:Cmu}) and~(\ref{eq:Cres}).  The relations we have between the 0-jet and 1-jet cross sections allow us to determine the covariance matrices $C_\mu$ and $C_\res$.  This has been discussed in detail in the previous section, where all entries in the matrices have been given explicitly.  In this section, we will give the numerical cross sections in each jet bin, their uncertainties, and the covariance matrices that will allow for the calculation of the uncertainties in any observable based on these rates.  We will compare the results of the resummed predictions to those obtained at fixed order, to demonstrate the improvement obtained.  We perform this study for the ATLAS and CMS parameters, taking $m_H = 125 \GeV$:
\be \label{eq:params}
\text{ATLAS: } \; \pTcut = 25 \GeV ,\, R = 0.4 \,, \qquad\qquad \text{CMS: } \; \pTcut = 30 \GeV ,\, R = 0.5 \,.
\ee
As in our previous numerical study we use MSTW NNLO PDFs, and assume an 8 TeV LHC.

\subsection{Results at Fixed Order}
\label{subsec:numericsFO}

We now present numerical results at fixed order using the ST method as discussed in \subsec{FO}.  Since this is how the uncertainties in the 0-jet and 1-jet bins are currently evaluated, the values we give here are intended to be benchmarks for comparison to our resummed results.  Recall that the cross sections for exclusive jet bins are obtained from differences of inclusive cross sections, meaning that we need inclusive cross sections with their scale variations.  These and the exclusive cross sections are listed in \tab{FOrates}.
\begin{table}[!htdp]
\begin{center}
\begin{tabular}{c||c|c}
\hline\hline
\; rate [pb]  \;&\; ATLAS ($\pTcut = 25 \GeV,\, R = 0.4$) \;&\; CMS ($\pTcut = 30 \GeV,\, R = 0.5$) \\ [0.3ex]
\hline\hline
$\sigma_\tot^{\textsc{nnlo}}$ \;&\; 19.27 $\pm$ 1.50 \;&\; 19.27 $\pm$ 1.50 \; \\
$\sigma_{\ge 1}^{\textsc{nlo}}$ \;&\; 7.85 $\pm$ 1.41 \;&\; 6.47 $\pm$ 1.27 \; \\
$\sigma_{\ge 2}^{\textsc{lo}}$ \;&\; 2.42 $\pm$ 1.80 \;&\; 1.73 $\pm$ 1.31 \; \\ [0.5ex] \hline
$\sigma_0^{\textsc{nnlo}}$ \;&\; 11.69 $\pm$ 2.06 \;&\; 12.80 $\pm$ 1.97 \; \\
$\sigma_1^{\textsc{nlo}}$ \;&\; 5.16 $\pm$ 2.29 \;&\; 4.75 $\pm$ 1.82 \; \\
\hline\hline
\end{tabular}
\end{center}
\caption{Fixed order cross sections and their uncertainties for ATLAS and CMS parameters.  The central scale is $\mu = m_H / 2$.}
\label{tab:FOrates}
\end{table}

We obtain the result for the total inclusive cross section from the Higgs cross section working group~\cite{LHCXS}, while the results for the inclusive 1-jet and 2-jet cross sections are obtained from MCFM.  As discussed previously, we use the central scale choice $\mu=m_H/2$ for these fixed-order quantities.  We find the following numerical values for the entries of the covariance matrices, in the basis of the 0-jet, 1-jet, and inclusive 2-jet cross sections: for ATLAS,
\bea \label{eq:covFOATLAS}
C^{\rm ATLAS}_\FO (\{\sigma_0, \sigma_1, \sigma_{\ge 2}\}) &=& \left(
\begin{array}{ccc}
1.50^2 + 1.41^2 & -1.41^2 & 0\\
-1.41^2 & 1.41^2+1.80^2 & -1.80^2\\
0 & -1.80^2 & 1.80^2
\end{array} \right) \pb^2 \nn \\
&=& \left(
\begin{array}{ccc}
4.24 & -1.99 & 0\\
-1.99 & 5.23 & -3.24\\
0 & -3.24 & 3.24
\end{array} \right) \pb^2 \,,
\eea
while for CMS,
\bea \label{eq:covFOCMS}
C^{\rm CMS}_\FO (\{\sigma_0, \sigma_1, \sigma_{\ge 2}\}) &=& \left(
\begin{array}{ccc}
3.86 & -1.61 & 0 \\
-1.61 & 3.33 & -1.72 \\
0 & -1.72 & 1.72
\end{array} \right) \pb^2 \,.
\eea
We note that the relative uncertainties on the exclusive 0-jet and 1-jet bins are $\pm 18\%$ and $\pm 44\%$ respectively for ATLAS, and $\pm 15\%$ and $\pm 38\%$ respectively for CMS.  The relative uncertainty on the sum of 0-jet and 1-jet bins is obtained by summing the upper $2 \times 2$ block of the covariance matrix, taking the square root, and then dividing by the sum of the cross sections given in \tab{FOrates}.  The numerical result is $\pm 14\%$ for ATLAS and $\pm 11\%$ for CMS.

\subsection{Results with Resummation}
\label{subsec:numericsresum}

When resummation is used to make predictions for the jet bins, the uncertainties are reduced compared to fixed order.  The same parameters as in \eq{params} are used, with the resummation schemes outlined in \sec{lowpTJ}.  The yield and migration uncertainties in the 0-jet bin are determined through scale variation, and are provided by the results in Ref.~\cite{Stewart:2013faa}.  The uncertainties in the 1-jet bin are given in \sec{lowpTJ}, as is the correlation between the 0-jet and 1-jet bins, allowing us to completely determine the covariance matrix.

\begin{table}[!htdp]
\begin{center}
\begin{tabular}{c||c|c}
\hline\hline
\; rate [pb]  \;&\; ATLAS ($\pTcut = 25 \GeV ,\, R = 0.4$) \;&\; CMS ($\pTcut = 30 \GeV ,\, R = 0.5$) \\
[0.3ex] \hline\hline
$\sigma_\tot$ \;&\; 21.69 $\pm$ 1.49 \;&\; 21.69 $\pm$ 1.49 \\ [0.3ex] \hline
$\sigma_0$ \;&\; 12.67 $\pm$ 0.87$_\mu$ $\pm$ 0.86$_\res$ ($\pm \; 1.22_\tot$) \;&\; 13.86 $\pm$ 0.70$_\mu$ $\pm$ 0.52$_\res$ ($\pm \; 0.87_\tot$) \; \\ [0.3ex] \hline
$\sigma_1$ \;&\; 5.68 $\pm$ 0.30$_\mu$ $\pm$ 0.89$_\res$ ($\pm \; 0.94_\tot$) \;&\; 4.97 $\pm$ 0.43$_\mu$ $\pm$ 0.61$_\res$ ($\pm \; 0.74_\tot$) \; \\ [0.3ex] \hline
$\sigma_{\ge 2}$ \;&\; 3.34 $\pm$ 0.32$_\mu$ $\pm$ 0.47$_\res$ ($\pm \; 0.57_\tot$) \;&\; 2.86 $\pm$ 0.36$_\mu$ $\pm$ 0.44$_\res$ ($\pm \; 0.57_\tot$) \; \\ [0.3ex]
\hline\hline
\end{tabular}
\end{center}
\caption{The total, 0-jet, 1-jet, and inclusive 2-jet cross sections and their yield, migration, and total uncertainties in Scheme A.}
\label{tab:resumrates}
\end{table}
In \tab{resumrates} we show the total, 0-jet, 1-jet, and inclusive 2-jet cross sections in Scheme A, along with their uncertainties.  These can be compared with the results in \tab{FOrates}.  As a comparison, the inclusive cross section and its uncertainty in Scheme B is $18.97 \pm 2.92$, the 0-jet cross sections are $12.44 \pm 1.59 \pb$ (ATLAS) and $13.29 \pm 1.63 \pb$ (CMS), the 1-jet cross sections are $3.70 \pm 1.13 \pb$ (ATLAS) and $3.26 \pm 0.93 \pb$ (CMS), and the inclusive 2-jet cross sections are $2.83 \pm 1.33 \pb$ (ATLAS) and $2.42 \pm 1.35 \pb$ (CMS).

The uncertainties in Scheme B improve on those at fixed order while the uncertainties in Scheme A are approximately a factor of two smaller than the fixed order ones.  We note that the cross section in Scheme A, our preferred matching scheme, is higher than the value obtained by the Higgs cross section working group~\cite{LHCXS}.  We believe that the inclusive cross section used in experimental analyses will eventually have to be updated to a value closer to the number found here, which accounts for newer perturbative information that has recently become available and provides a better matching between the best predictions for the 0-jet and 1-jet bins.  Not only does the effective-field theory framework used here give a value higher than the current result, recent approximate N$^3$LO evaluations of the cross section~\cite{Ball:2013bra} that feature an improved treatment of the singular limits find a cross section in better agreement with our Scheme A value.  As a temporary way of rendering our analysis consistent with the currently-assumed total cross section, we advocate a global rescaling of the Scheme A numbers in \tab{resumrates} with the ratio of the Higgs cross section working group cross section over our $\sigma_\tot$.

\begin{figure}[!t]
\begin{center}
\includegraphics[width=0.48\textwidth]{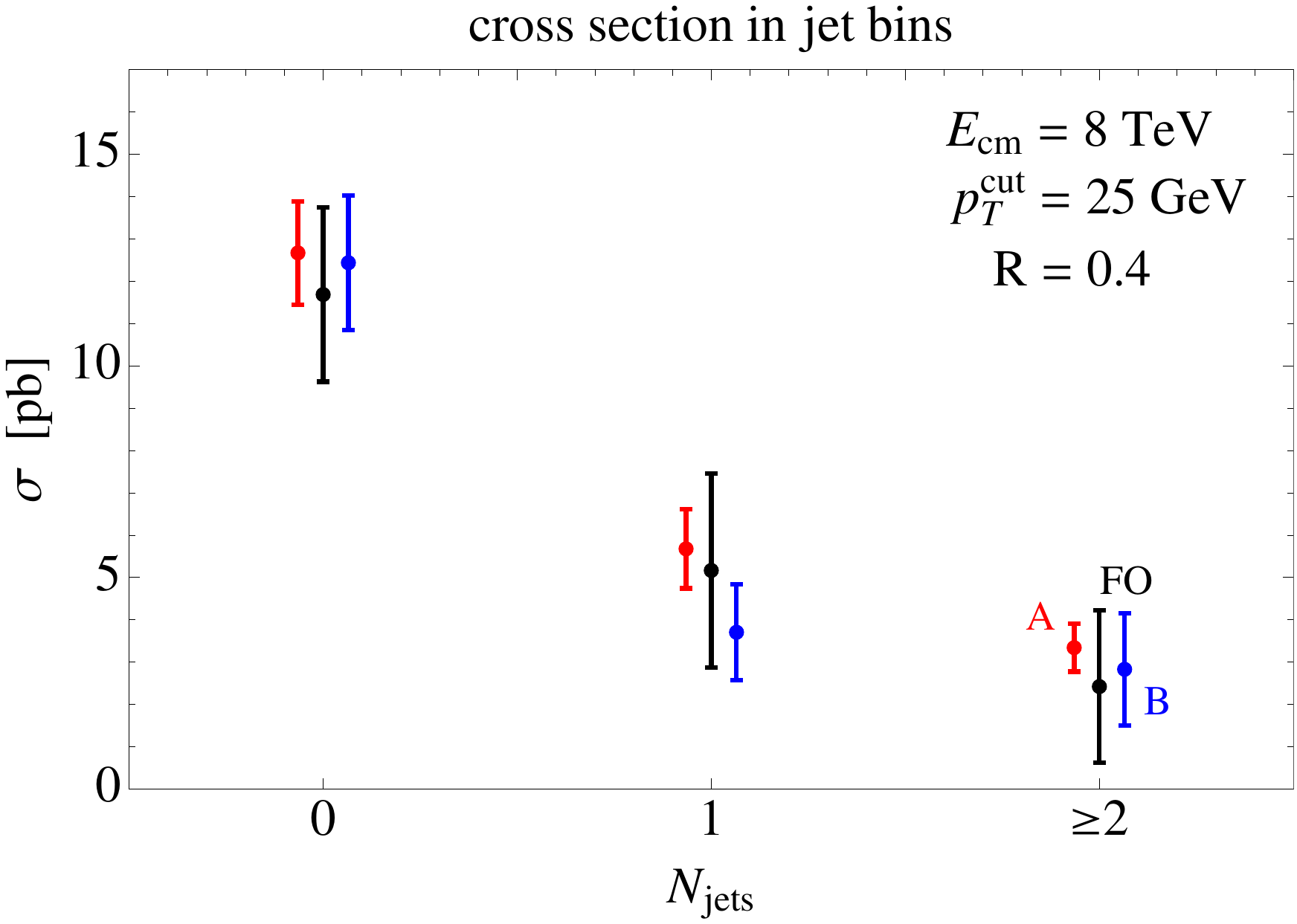} \hspace{2ex}
\includegraphics[width=0.48\textwidth]{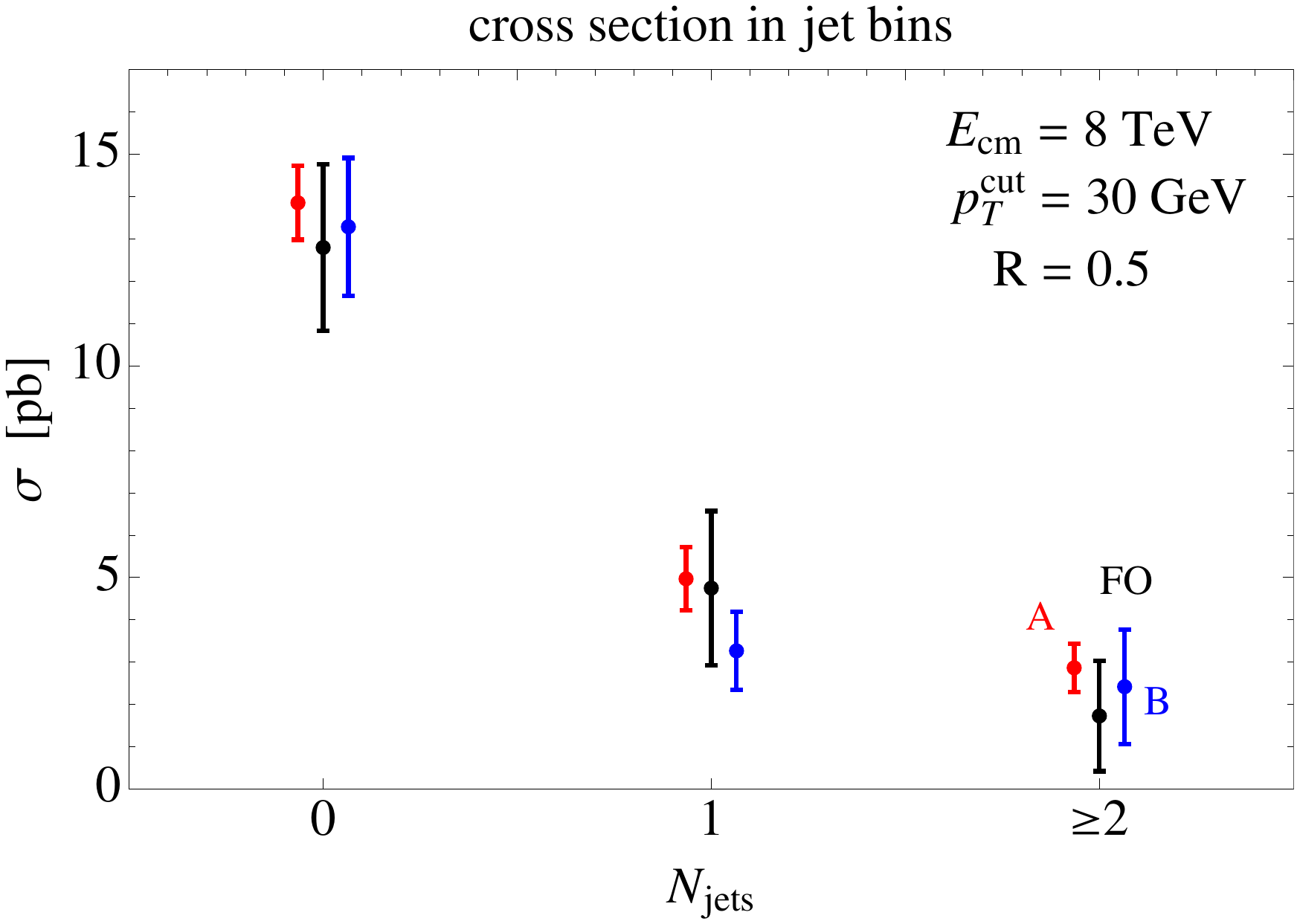}
\end{center}
\vspace{-0.5cm}
\caption{The cross sections in the 0-jet, 1-jet, and inclusive 2-jet bins, for ATLAS (left) and CMS (right) parameter choices.  In each jet bin, Scheme A (red left points), Scheme B (blue, right points) are compared to the fixed-order results (black, middle points).} \label{fig:jetrates}
\end{figure}

In \fig{jetrates}, we plot the cross sections in each jet bin against the fixed-order cross sections.  The agreement between fixed order and the two matching schemes, as well as the reduced uncertainties when moving beyond fixed order, are clearly visible in the plot.  We note that these results, weighted by experimental efficiencies in each jet bin, may directly be used in analyses of the detailed properties of Higgs events. In particular, they can be used to compare to the recent ATLAS $H\to\gamma\gamma$ measurements of fiducial jet-bin cross sections~\cite{ATLAS-CONF-2013-072}, after taking into account experimental efficiencies and the $H\to\gamma\gamma$ branching ratio.

Finally, we give the covariance matrices for matching Scheme A: for ATLAS,
\begin{align} \label{eq:covAATLAS}
C^{\rm ATLAS} = C^{\rm ATLAS}_\mu + C^{\rm ATLAS}_\res &= \left(
\begin{array}{ccc}
0.75 & 0.26 & 0.28 \\
0.26 & 0.09 & 0.10 \\
0.09 & 0.10 & 0.10
\end{array} \right) \pb^2 + \left(
\begin{array}{ccc}
0.74 & -0.65 & -0.08 \\
-0.65 & 0.79 & -0.13 \\
-0.08 & -0.13 & 0.22
\end{array} \right) \pb^2 \nn \\
&= \left(
\begin{array}{ccc}
1.49 & -0.39 & 0.20 \\
-0.39 & 0.88 & -0.04 \\
0.20 & -0.04 & 0.32
\end{array} \right) \pb^2 \,,
\end{align}
and for CMS,
\begin{align} \label{eq:covACMS}
C^{\rm CMS} = C^{\rm CMS}_\mu + C^{\rm CMS}_\res &= \left(
\begin{array}{ccc}
0.50 & 0.30 & 0.25 \\
0.30 & 0.18 & 0.15 \\
0.25 & 0.15 & 0.13
\end{array} \right) \pb^2 + \left(
\begin{array}{ccc}
0.27 & -0.22 & -0.05 \\
-0.22 & 0.37 & -0.15 \\
-0.05 & -0.15 & 0.19
\end{array} \right) \pb^2 \nn \\
&= \left(
\begin{array}{ccc}
0.76 & 0.09 & 0.20 \\
0.09 & 0.55 & 0.01 \\
0.20 & 0.01 & 0.32
\end{array} \right) \pb^2 \,.
\end{align}
Like the fixed-order covariance matrices in \eqs{covFOATLAS}{covFOCMS}, these results may be used to determine the correlations in any observable built from the 0-jet, 1-jet, and inclusive 2-jet cross sections. The relative uncertainties on the exclusive 0-jet and 1-jet bins have decreased to $\pm 10\%$ and $\pm 16\%$ respectively for ATLAS and $\pm 6\%$ and $\pm15\%$ respectively for CMS, while the numerical uncertainty on the sum of 0-jet and 1-jet bins is now $\pm 7\%$ for ATLAS and $\pm 6\%$ for CMS.

\subsection{Impact on the Determination of the Signal Strength}
\label{subsec:numericsmu}

A key measurement in every given production/decay channel is the coupling of the Higgs boson to other Standard Model particles.  In the simplest case, this corresponds to a measurement of the signal strength $\mu$, the ratio of the experimentally observed cross section to the SM expectation.  For the $H\to WW^*$ decay channel the theoretical uncertainties discussed here, which are those for the expected signal in the sum of jet bins, are a major component of the theory systematics on $\mu$.  Using the cross sections and the covariance matrices given above, we can estimate the reduction in theoretical uncertainties on $\mu$ that is gained by using resummed results for the cross sections.

The signal strength is defined as the ratio of observed and expected cross sections that pass a set of analysis cuts,
\be
\mu = \frac{\sigma^{\rm obs}_{\rm acc.}}{\sigma^{\rm exp}_{\rm acc.}} \,.
\ee
ssThe relative theory uncertainty in the signal strength is then equal to the relative theory uncertainty of the expected signal cross section:
\be
\frac{\Delta \mu}{\mu} = \frac{\Delta \sigma^{\rm exp}_{\rm acc.}}{\sigma^{\rm exp}_{\rm acc.}} \,.
\ee
These expected and observed rates can be written as the sum of cross sections in each jet bin, with acceptances that depend on the cuts in each bin.  For the expected rate,
\be
\sigma^{\rm exp}_{\rm acc.} = \epsilon_0^{\rm exp} \sigma_0 (\pTcut) + \epsilon_1^{\rm exp} \sigma_1 ([\pTcut, \infty]; \pTcut) + \epsilon_{\ge 2}^{\rm exp} \sigma_{\ge 2} (\pTcut, \pTcut) \,.
\ee
The efficiencies $\epsilon_i^{\rm exp}$ are particular to a given analysis, and include the effects of both kinematic selection cuts in the analysis (e.g. cuts on the leptonic final state) as well as all experimental efficiencies (such as from trigger, reconstruction, and particle identification).

Because the measurement sums over contributions from different numbers of jets, the final perturbative uncertainty is reduced compared to that of the individual jet bins due to the anticorrelations between the bins.  For the same reason the relative uncertainty on the total cross section is lower than the relative uncertainty in individual jet bins.  If the efficiencies $\epsilon_i^{\rm exp}$ in each jet bin are the same, then the uncertainty squared is proportional to the sum over all elements of the covariance matrix, which cancels the migration effects.  If the measurement is dominated by the 0-jet and 1-jet bins (as is the case for gluon fusion), then the sum of entries in the 0-jet and 1-jet block of the covariance matrix determines the uncertainty.  For the fixed-order and resummed results above, this implies that the relative uncertainty in the combined (0+1)-jet cross section is 6.9\% for ATLAS and 6.5\% for CMS using the resummed results and 13.9\% for ATLAS and 11.3\% for CMS using fixed-order results.

Of course, in general the 0-jet and 1-jet bins will have different acceptances for a given analysis.  As an illustrative example, we use the ATLAS $H \to WW^*$ analysis results to estimate the uncertainty on $\mu$ for the resummed results~\cite{ATLAS:2013wla}.  We also perform the analysis using the fixed-order cross section predictions, which allows us to compare the uncertainty in $\mu$ against the value determined by ATLAS (which uses the fixed-order results).  We expect that the results of this exercise will be similar for the CMS $H\to WW^*$ analysis~\cite{CMS:bxa}.

The $H \to WW^*$ analysis uses leptonic final states with electrons and muons, and divides these final states into two channels: those with two electrons or two muons ($ee + \mu\mu$), and those with one electron and one muon ($e\mu + \mu e$).  The cuts in each channel are different, with more stringent cuts in the $ee+\mu\mu$ channel to suppress the Drell-Yan background.  Therefore, the efficiency in jet bin $j$ is
\begin{align}
\epsilon^{\rm exp}_j &= \epsilon (ee+\mu\mu \text{ channel, jet bin } j) \, \epsilon_j^{ee+\mu\mu}+ \epsilon (e\mu+\mu e \text{ channel, jet bin } j) \, \epsilon_j^{e\mu+\mu e} \,,
\end{align}
with
\begin{align}
&\epsilon (ee+\mu\mu \text{ channel, jet bin } j) \, : \, \text{efficiency to be in the $ee+\mu\mu$ channel for jet bin $j$} \,, \nn \\
&\epsilon_j^{ee+\mu\mu} \, : \, \text{efficiency of additional selection cuts in the $ee+\mu\mu$ channel for jet bin $j$} \,,
\end{align}
and analogously for the $e\mu + \mu e$ channel.  The efficiencies to be in the $ee+\mu\mu$ or $e\mu + \mu e$ channels include the branching ratio for the $W$ decays into the allowed final states as well as the selection cuts for the channel.  The efficiencies for the same leptonic channel can differ between jet bins, since the selection cuts depend on the final state lepton kinematics (which change if there are high $p_T$ jets in the final state, for example in the 1-jet bin).

The efficiencies $\epsilon_j^{ee+\mu\mu}$ and $\epsilon_j^{e\mu+\mu e}$ for each jet bin can be extracted from Tables 8--10 in Ref.~\cite{ATLAS:2013wla}.  They are
\begin{align}
\epsilon_0^{ee+\mu\mu} &= 0.40 \,, \!\!\!\!\!\!\!\!\!\!\!\!\!\!\!\!&\!\!\!\!\!\!\!\!\!\!\!\!\!\!\!\! \epsilon_0^{e\mu+\mu e} &= 0.69 \,, \nn \\
\epsilon_1^{ee+\mu\mu} &= 0.30 \,, \!\!\!\!\!\!\!\!\!\!\!\!\!\!\!\!&\!\!\!\!\!\!\!\!\!\!\!\!\!\!\!\! \epsilon_1^{e\mu+\mu e} &= 0.61 \,, \nn \\
\epsilon_{\ge 2}^{ee+\mu\mu} &= 0.02 \,, \!\!\!\!\!\!\!\!\!\!\!\!\!\!\!\!&\!\!\!\!\!\!\!\!\!\!\!\!\!\!\!\! \epsilon_{\ge 2}^{e\mu+\mu e} &= 0.02 \,.
\end{align}
The analysis cuts in the inclusive 2-jet bin are targeted to the VBF topology, causing the $gg$ fusion efficiency to be small.  Because the inclusive 2-jet cross section is already relatively small, we neglect this bin in the remainder of our analysis.

The efficiencies for the various leptonic channels in the jet bins are not available, but the ratios between channels in each jet bin are
\begin{align} \label{eq:lepratios}
\frac{\epsilon (ee+\mu\mu \text{ channel, 0-jet bin})}{\epsilon (e\mu+\mu e \text{ channel, 0-jet bin})} &= 0.60 \,, \nn \\
\frac{\epsilon (ee+\mu\mu \text{ channel, 1-jet bin})}{\epsilon (e\mu+\mu e \text{ channel, 1-jet bin})} &= 0.56 \,.
\end{align}
Making a mild assumption, we will use these ratios to determine the relative uncertainty on $\sigma^{\rm exp}_{\rm acc.}$.  We will assume
\be \label{eq:effassumption}
\epsilon (e\mu+\mu e \text{ channel, 0-jet bin}) = \epsilon (e\mu+\mu e \text{ channel, 1-jet bin}) \,,
\ee
and we make this assumption for the $e\mu + \mu e$ channel instead of the $ee+\mu\mu$ channel because the selection cuts are stricter in the $ee+\mu\mu$ channel.  This assumption is supported by the fact that the ratios in \eq{lepratios} are very similar.

Extracting the common efficiency in \eq{effassumption}, the efficiency for each jet bin is
\begin{align}
\epsilon^{\rm exp}_j &= \epsilon (e\mu+\mu e \text{ channel, jet bin } j) \biggl[ \frac{\epsilon (ee+\mu\mu \text{ channel, jet bin } j)}{\epsilon (e\mu+\mu e \text{ channel, jet bin } j)}\, \epsilon_j^{ee+\mu\mu} + \epsilon_j^{e\mu+\mu e} \biggr] \nn \\
&\approx \epsilon (e\mu+\mu e \text{ channel}) \biggl[ \frac{\epsilon (ee+\mu\mu \text{ channel, jet bin } j)}{\epsilon (e\mu+\mu e \text{ channel, jet bin } j)}\, \epsilon_j^{ee+\mu\mu} + \epsilon_j^{e\mu+\mu e} \biggr] \,.
\end{align}
The overall efficiency $\epsilon (e\mu+\mu e \text{ channel})$ is irrelevant for the relative uncertainty of $\sigma^{\rm exp}_{\rm acc.}$.  Putting the numbers together, the relative uncertainty is
\begin{align} \label{eq:uncertmu}
\text{fixed order} &: \; \frac{\Delta \mu}{\mu} = 13.3\% \,, \nn \\
\text{Scheme A} &: \; \frac{\Delta \mu}{\mu} = 6.9\% \,.
\end{align}
We note that making the assumption that the $\epsilon (ee+\mu\mu \text{ channel})$ is equal for all jet bins, the relative uncertainty in $\mu$ changes by less than $0.5\%$ for either case.  The ATLAS measurement, using fixed-order results for the cross sections and uncertainties, finds the relative uncertainty on $\mu$ from the expected signal to be $+12\%$ and $-9\%$, giving confidence to our estimates.  The result in \eq{uncertmu} is encouraging, as it shows that the theoretical uncertainties in the expected signal in the determination of $\mu$ are approximately halved by using the results in Scheme A.

Note that in the above example we assumed for simplicity that the signal-to-background ratio is the same in the 0-jet and 1-jet bins.  In general, they can be different, giving different weights to each bin in the final combination.  This will cause the migration uncertainties to only partially cancel (as in the case where the signal efficiencies in each jet bin are different), and can lead to an increase or decrease in the uncertainty in the (0+1)-jet combination.  Of course, the resummed predictions provide an improvement compared to fixed order irrespective of whether one jet bin dominates the combination or both contribute equally, since the resummation reduces the uncertainties in each individual jet bin.

\section{Conclusions}
\label{sec:conclusions}

In this work we have studied the theoretical uncertainties affecting the $H+0$-jet, $H+1$-jet, and inclusive $H+2$-jet cross sections.  We have presented a complete method for turning the separate resummations of the large logarithms in the 0-jet and 1-jet bins into a combined covariance matrix for use in experimental studies.  A key element of our approach is an extension of the $H+1$-jet resummation into the phase space region where the transverse momentum of the final state jet is small.  This is the first result that resums the exclusive 1-jet cross section across the entire final state phase space. Our approach provides a systematically-improvable framework that can incorporate improved calculations as they become available.  As a demonstration, we have shown how adding enhanced corrections in the 0-jet and 1-jet bins leads to both a smoother matching between the two bins and reduced perturbative uncertainties.  Future calculations, such as the NNLO fixed-order result for $H+1$-jet, are easily included in our approach.

For phenomenologically relevant parameters, we have given the fixed-order and resummed predictions for the cross sections and their uncertainties in \sec{numerics}. Our results can be used to compare to data in more detailed Higgs cross section measurements, such as the ATLAS $H\to\gamma\gamma$ differential cross section analysis. As an example, we have used our results to estimate the impact on the theoretical uncertainties on the signal strength in the $H\to WW^*$ channel.  We find nearly a factor of two reduction in the theoretical uncertainty on the signal strength.  This is a dramatic improvement in the estimated uncertainty, and will become crucial as the statistical uncertainties on Higgs cross sections continue to decrease during Run II of the LHC.

Although we have focused on the $WW^{*}$ final state, we expect the improvements we find to be applicable to other Higgs analyses in which the inclusive 2-jet bin is treated separately from the 0-jet and 1-jet bins.  In \subsec{numericsmu}, we have discussed the general improvement that can be expected with our results, finding that the uncertainties in the combined resummed (0+1)-jet cross sections, with no experimental acceptance cuts, are approximately half those at fixed order.  This is consistent with the uncertainty reduction that we find in \subsec{numericsmu} for the study specific to the ATLAS $H\to WW^*$ analysis.

The same machinery that we have developed here may be used to study exclusive jet cross sections for other processes.  Drell-Yan and $W$ production are notable examples, where precise measurements of the exclusive jet cross sections are available, and may be used to test this framework.  We look forward to the application of our ideas in future LHC analyses.

\section{Acknowledgments}
\label{sec:acks}

We thank Joey Huston for helpful discussions.  JW thanks Northwestern University for hospitality while portions of this work were completed.

The work of RB was supported by the U.S. Department of Energy, Division of High Energy Physics, under contract DE-AC02-06CH11357.  The work of XL and FP was supported by the U.S. Department of Energy, Division of High Energy Physics, under contract DE-AC02-06CH11357 and the grants DE-FG02-95ER40896 and DE-FG02-08ER4153.  The work of FT was supported by the DFG Emmy-Noether Grant No. TA 867/1-1.  The work of JW was supported by the Director, Office of Science, Office of High Energy Physics of the U.S. Department of Energy under the Contract No. DE-AC02-05CH11231.


\begin{thebibliography}{99}

\bibitem{Aad:2012tfa} 
  G.~Aad {\it et al.}  [ATLAS Collaboration],
  Phys.\ Lett.\ B {\bf 716}, 1 (2012)
  [arXiv:1207.7214 [hep-ex]].

\bibitem{Aad:2012uub} 
  G.~Aad {\it et al.}  [ATLAS Collaboration],
  Phys.\ Lett.\ B {\bf 716}, 62 (2012)
  [arXiv:1206.0756 [hep-ex]].

\bibitem{Chatrchyan:2012ufa} 
  S.~Chatrchyan {\it et al.}  [CMS Collaboration],
  Phys.\ Lett.\ B {\bf 716}, 30 (2012)
  [arXiv:1207.7235 [hep-ex]].

\bibitem{Chatrchyan:2012ty} 
  S.~Chatrchyan {\it et al.}  [CMS Collaboration],
  Phys.\ Lett.\ B {\bf 710}, 91 (2012)
  [arXiv:1202.1489 [hep-ex]].

\bibitem{Cacciari:2008gp} 
  M.~Cacciari, G.~P.~Salam and G.~Soyez,
  JHEP {\bf 0804}, 063 (2008)
  [arXiv:0802.1189 [hep-ph]].
  
\bibitem{Stewart:2011cf} 
  I.~W.~Stewart and F.~J.~Tackmann,
  Phys.\ Rev.\ D {\bf 85}, 034011 (2012)
  [arXiv:1107.2117 [hep-ph]].

\bibitem{Gangal:2013nxa}
  S.~Gangal and F.~J.~Tackmann,
  Phys.\ Rev.\ D {\bf 87}, 093008 (2013)
  [arXiv:1302.5437 [hep-ph]].

\bibitem{Stewart:2013faa} 
  I.~W.~Stewart, F.~J.~Tackmann, J.~R.~Walsh and S.~Zuberi,
  arXiv:1307.1808 [hep-ph].

\bibitem{Anastasiou:2005qj} 
  C.~Anastasiou, K.~Melnikov and F.~Petriello,
  Nucl.\ Phys.\ B {\bf 724}, 197 (2005)
  [hep-ph/0501130].

\bibitem{Catani:2007vq} 
  S.~Catani and M.~Grazzini,
  Phys.\ Rev.\ Lett.\  {\bf 98}, 222002 (2007)
  [hep-ph/0703012].

\bibitem{Glosser:2002gm} 
  C.~J.~Glosser and C.~R.~Schmidt,
  JHEP {\bf 0212}, 016 (2002)
  [hep-ph/0209248].

\bibitem{Ravindran:2002dc} 
  V.~Ravindran, J.~Smith and W.~L.~Van Neerven,
  Nucl.\ Phys.\ B {\bf 634}, 247 (2002)
  [hep-ph/0201114].

\bibitem{Czakon:2010td} 
  M.~Czakon,
  Phys.\ Lett.\ B {\bf 693}, 259 (2010)
  [arXiv:1005.0274 [hep-ph]].

\bibitem{Boughezal:2011jf} 
  R.~Boughezal, K.~Melnikov and F.~Petriello,
  Phys.\ Rev.\ D {\bf 85}, 034025 (2012)
  [arXiv:1111.7041 [hep-ph]].

\bibitem{Boughezal:2013uia} 
  R.~Boughezal, F.~Caola, K.~Melnikov, F.~Petriello and M.~Schulze,
  JHEP {\bf 1306}, 072 (2013)
  [arXiv:1302.6216 [hep-ph]].

\bibitem{Aglietti:2004nj} 
  U.~Aglietti, R.~Bonciani, G.~Degrassi and A.~Vicini,
  Phys.\ Lett.\ B {\bf 595}, 432 (2004)
  [hep-ph/0404071].
  
  \bibitem{Actis:2008ug} 
  S.~Actis, G.~Passarino, C.~Sturm and S.~Uccirati,
  Phys.\ Lett.\ B {\bf 670}, 12 (2008)
  [arXiv:0809.1301 [hep-ph]].
  
  \bibitem{Anastasiou:2008tj} 
  C.~Anastasiou, R.~Boughezal and F.~Petriello,
  JHEP {\bf 0904}, 003 (2009)
  [arXiv:0811.3458 [hep-ph]].
  
  \bibitem{Keung:2009bs} 
  W.~-Y.~Keung and F.~J.~Petriello,
  Phys.\ Rev.\ D {\bf 80}, 013007 (2009)
  [arXiv:0905.2775 [hep-ph]].
  
  \bibitem{Brein:2010xj} 
  O.~Brein,
  Phys.\ Rev.\ D {\bf 81}, 093006 (2010)
  [arXiv:1003.4438 [hep-ph]].
  
  \bibitem{Passarino:2013nka} 
  G.~Passarino,
  arXiv:1308.0422 [hep-ph].

\bibitem{Spira:1995rr} 
  M.~Spira, A.~Djouadi, D.~Graudenz and P.~M.~Zerwas,
  Nucl.\ Phys.\ B {\bf 453}, 17 (1995)
  [hep-ph/9504378].
  
  \bibitem{Harlander:2009my} 
  R.~V.~Harlander, H.~Mantler, S.~Marzani and K.~J.~Ozeren,
  Eur.\ Phys.\ J.\ C {\bf 66}, 359 (2010)
  [arXiv:0912.2104 [hep-ph]].
  
  \bibitem{Harlander:2012hf} 
  R.~V.~Harlander, T.~Neumann, K.~J.~Ozeren and M.~Wiesemann,
  JHEP {\bf 1208}, 139 (2012)
  [arXiv:1206.0157 [hep-ph]].

\bibitem{Banfi:2012yh} 
  A.~Banfi, G.~P.~Salam and G.~Zanderighi,
  JHEP {\bf 1206}, 159 (2012)
  [arXiv:1203.5773 [hep-ph]].
  
  \bibitem{Banfi:2012jm} 
  A.~Banfi, P.~F.~Monni, G.~P.~Salam and G.~Zanderighi,
  Phys.\ Rev.\ Lett.\  {\bf 109}, 202001 (2012)
  [arXiv:1206.4998 [hep-ph]].
  
  \bibitem{Becher:2012qa} 
  T.~Becher and M.~Neubert,
  JHEP {\bf 1207}, 108 (2012)
  [arXiv:1205.3806 [hep-ph]].
  
\bibitem{Banfi:2013eda} 
  A.~Banfi, P.~F.~Monni and G.~Zanderighi,
  arXiv:1308.4634 [hep-ph].
  
  \bibitem{Becher:2013xia} 
  T.~Becher, M.~Neubert and L.~Rothen,
  JHEP {\bf 1310}, 125 (2013)
  [arXiv:1307.0025 [hep-ph]].
 
 \bibitem{Tackmann:2012bt} 
  F.~J.~Tackmann, J.~R.~Walsh and S.~Zuberi,
  Phys.\ Rev.\ D {\bf 86}, 053011 (2012)
  [arXiv:1206.4312 [hep-ph]].
  
  \bibitem{Liu:2012sz} 
  X.~Liu and F.~Petriello,
  Phys.\ Rev.\ D {\bf 87}, 014018 (2013)
  [arXiv:1210.1906 [hep-ph]].

\bibitem{Liu:2013hba} 
  X.~Liu and F.~Petriello,
  Phys.\ Rev.\ D {\bf 87}, 094027 (2013)
  [arXiv:1303.4405 [hep-ph]].

\bibitem{Dasgupta:2001sh} 
  M.~Dasgupta and G.~P.~Salam,
  Phys.\ Lett.\ B {\bf 512}, 323 (2001)
  [hep-ph/0104277].

\bibitem{Campbell:2010ff} 
  J.~M.~Campbell and R.~K.~Ellis,
  Nucl.\ Phys.\ Proc.\ Suppl.\  {\bf 205-206}, 10 (2010)
  [arXiv:1007.3492 [hep-ph]].

\bibitem{Campbell:2010cz}
  J.~M.~Campbell, R.~K.~Ellis and C.~Williams,
  Phys.\ Rev.\ D {\bf 81}, 074023 (2010)
  [arXiv:1001.4495 [hep-ph]].

\bibitem{Dittmaier:2011ti} 
  S.~Dittmaier {\it et al.}  [LHC Higgs Cross Section Working Group Collaboration],
  arXiv:1101.0593 [hep-ph].

\bibitem{Martin:2009iq} 
  A.~D.~Martin, W.~J.~Stirling, R.~S.~Thorne and G.~Watt,
  Eur.\ Phys.\ J.\ C {\bf 63}, 189 (2009)
  [arXiv:0901.0002 [hep-ph]].

\bibitem{Parisi:1979xd}
  G.~Parisi,
  Phys.\ Lett.\ B {\bf 90}, 295 (1980).

\bibitem{Sterman:1986aj}
  G.~F.~Sterman,
  Nucl.\ Phys.\ B {\bf 281}, 310 (1987).

\bibitem{Magnea:1990zb}
  L.~Magnea and G.~F.~Sterman,
  Phys.\ Rev.\ D {\bf 42}, 4222 (1990).

\bibitem{Ahrens:2008qu} 
  V.~Ahrens, T.~Becher, M.~Neubert and L.~L.~Yang,
  Phys.\ Rev.\ D {\bf 79}, 033013 (2009)
  [arXiv:0808.3008 [hep-ph]].

\bibitem{Ahrens:2008nc} 
  V.~Ahrens, T.~Becher, M.~Neubert and L.~L.~Yang,
  Eur.\ Phys.\ J.\ C {\bf 62}, 333 (2009)
  [arXiv:0809.4283 [hep-ph]].

\bibitem{Gehrmann:2011aa} 
  T.~Gehrmann, M.~Jaquier, E.~W.~N.~Glover and A.~Koukoutsakis,
  JHEP {\bf 1202}, 056 (2012)
  [arXiv:1112.3554 [hep-ph]].

\bibitem{LHCXS}
\url{https://twiki.cern.ch/twiki/bin/view/LHCPhysics/CrossSections}      

\bibitem{Ball:2013bra} 
  R.~D.~Ball, M.~Bonvini, S.~Forte, S.~Marzani and G.~Ridolfi,
  Nucl.\ Phys.\ B {\bf 874}, 746 (2013)
  [arXiv:1303.3590 [hep-ph]].

\bibitem{ATLAS-CONF-2013-072}
   ATLAS Collaboaration,
   ATLAS-CONF-2013-072.

\bibitem{ATLAS:2013wla} 
  ATLAS Collaboration,
  ATLAS-CONF-2013-030.

\bibitem{CMS:bxa} 
  CMS Collaboration,
  CMS-PAS-HIG-13-003.

\end{thebibliography}
\end{document}